\documentclass[12pt]{article}
\usepackage{here,amsfonts,amssymb,pstricks,pst-node,
pst-plot,pst-coil,epsfig,psfig,graphicx}
\usepackage{amsmath,xy}
\xyoption{all}

\newbox\sample
\newif\ifproofmode
\newif\ifsymindex
\global\symindexfalse
\newwrite\inx
\def\indsyma#1#2{\ifproofmode\marginpar{$\scriptstyle#1$}\fi%
\ifx#2\empty\write\inx{$\noexpand#1$,\space\thepage}%
\write\inx{\string\newline}\else%
\write\inx{$\noexpand#1$,\space#2,\space\thepage}%
\write\inx{\string\newline}\fi\ignorespaces}%
\def\indsym#1#2{\ifsymindex%
\ifproofmode\marginpar{$\scriptstyle#1$}\fi%
\ifx#2\empty\write\inx{\string\item \space$\noexpand#1$,\space\thepage}%
\else%
\write\inx{\string\item \space$\noexpand#1$,\space#2,\space\thepage}%
\fi\ignorespaces\fi}%
\newskip\dangerskipb
\newskip\dangerskip
\def\hang{\hangindent\dangerskip}

\def\s#1{{\cal #1}}

\def\lag{\left\langle}
\def\rag{\right\rangle}

\def\pairt#1#2{\lag #1, #2\rag}

\def\proof{\noindent{\it Proof\/}.\enspace}

\def\remark{\bigskip\noindent{\bf Remark:}\enspace}

\font\manual=manfnt at 12pt
\def\danbend{{\manual\char127}}
\def\ddatanger{\medbreak\begingroup\clubpenalty=10000
 \def\par{\endgraf\endgroup\medbreak} \noindent\hang\hangafter=-2
 \hbox to0pt{\hskip-3.5pc\danbend\kern1pt%
\danbend\hfill}}

\def\dobdownarrow{\mathop{\vbox{\kern2pt \hbox{$\Big\downarrow$}\kern-16.5pt
                          \nointerlineskip\hbox{$\Big\downarrow$}}}}

\def\lrightarrow{\hbox to 25pt{\rightarrowfill}}

\def\supexp{exp(m,n,p)=m^{m^{m^{\cdot^{\cdot^{\cdot^{m^{p}}}}}}}
\vbox{\hbox{$\Big\}\scriptstyle n$}\kern0pt}}

\def\supexpo#1#2#3{#1^{#1^{\cdot^{\cdot^{\cdot^{#1^{#2}}}}}}
\vbox{\hbox{$\Big\}\scriptstyle #3$}\kern0pt}}

\def\sqr#1#2{{\vcenter{\hrule height .#2pt
         \hbox{\vrule width.#2pt height#1pt \kern#1pt
             \vrule width.#2pt}
         \hrule height.#2pt}}}

\def\square{\mathchoice\sqr64\sqr64\sqr{2.1}3\sqr{1.5}3}

\def\bigsquare{\mathchoice\sqr76\sqr76\sqr{2.1}3\sqr{1.5}3}

\def\lag{\langle}
\def\rag{\rangle}

\def\co{\colon}

\newskip\bogcentering \bogcentering= 0pt plus 1000pt minus 1000pt 

\def\matth{\mathsurround=0pt}
\def\fakrightarrowfill{$\matth \mathord- \mkern-6mu
  \cleaders\hbox{$\mkern-2mu \mathord- \mkern-2mu$}\hfill
 \mkern-6mu \mathord\rightarrow$}

\def\fakoverrightarrow#1{\vbox{\ialign{##\crcr
  \fakrightarrowfill\crcr\noalign{\kern-1pt\nointerlineskip}
 $\hfil\displaystyle{#1}\hfil$\crcr}}}

\newif\ifdtatp

\def\displaty{%
\global \dtatptrue \openup \jot \matth \everycr{\noalign{\ifdtatp \global 
\dtatpfalse \vskip -\lineskiplimit \vskip \normallineskiplimit \else 
\penalty \interdisplaylinepenalty \fi }}}

\def\displaylignes#1{\displaty
   \halign{\hbox to\displaywidth{$\displaystyle##$}\crcr
   #1\crcr}}
\def\eqaligneno#1{\displaty \tabskip=\bogcentering
 \halign to\displaywidth{\hfil$\displaystyle{##}$\tabskip=0pt
 &$\displaystyle{{}##}$\hfil\tabskip=\bogcentering
 &\llap{$##$}\tabskip=0pt\crcr
 #1\crcr}}
\def\leqaligneno#1{\displaty \tabskip=\bogcentering
 \halign to\displaywidth{\hfil$\displaystyle{##}$\tabskip=0pt
 &$\displaystyle{{}##}$\hfil\tabskip=\bogcentering
 &\kern-\displaywidth\rlap{$##$}\tabskip=\displaywidthpt\crcr
 #1\crcr}}
\def\ligne{\hbox to\hsize}
\newdimen\nouvpagewidth
\newdimen\offwidth
\newdimen\lawidthoui
\nouvpagewidth=\lawidthoui
\def\kboxit#1{\vbox{\hrule\hbox{\vrule\kern3pt
              \vbox{\kern3pt#1\kern3pt}\kern3pt\vrule}\hrule}}

\def\kboxitb#1{\vbox{\hrule\hbox{\vrule\kern3pt
              \vbox{\kern3pt#1\kern3pt}\kern3pt\vrule}\hrule}}

\def\laboxaround#1{
\aboxaround{\hbox to\hsize{\hfill\box2\hfill}}{#1}
}

\def\boxar#1#2{
\aboxaround{\hbox to\hsize{\hfill#1\hfill}}{#2}
}

\def\aboxaround#1#2{
\setbox4=\vbox{\hsize #2\noindent\strut#1\strut}
\kboxitb{\box4}}

\def\kframeit#1{\vbox{\hrule\hbox{\vrule\kern5pt
              \vbox{\kern5pt#1\kern5pt}\kern5pt\vrule}\hrule}}
\newskip\savnormalbaselineskip
\newskip\savnormallineskip
\newdimen\savnormallineskiplimit

\def\square{\bigsquare}

\newtheorem{thm}{Theorem}[section]
\newtheorem{lemma}[thm]{Lemma}

\newtheorem{defin}[thm]{Definition}
\def\mdeg{m}
\def\ndeg{m}
\def\pdeg{p}
\def\qdeg{q}
\def\reals{\mathbb{R}}
\def\rprospac#1{\mathbb{RP}^{#1}}
\def\affreal{\reals}
\def\projs#1{\mathbb{P}(#1)}
\def\mapdef#1#2#3{#1\co #2\rightarrow #3}
\def\ptb#1{#1}
\def\hli#1{\widehat{#1}}
\def\vectorr#1{\overrightarrow{#1}}
\def\pcompl#1{\widetilde{#1}}
\def\affs{\s{E}}
\begin{document}
\title{Simple Methods For Drawing Rational Surfaces as
Four or Six  B\'ezier Patches}
\author{Jean Gallier\\
%author{Eugenio Bertini, Elie Cartan,\\
%and Kurt W.A.J.H.Y. Reillag\\
 \\
Department of Computer and Information Science\\
University of Pennsylvania\\
Philadelphia, PA 19104, USA\\ \\
%{\tt Bertini@pisa.mozarella.edu}\\
%{\tt Cartan@dolomieu.reblochon.edu}\\
{\tt jean@saul.cis.upenn.edu}\\
%{\tt dmd@gradient.cis.upenn.edu}
}
\maketitle
%\vspace{0.07cm}
\noindent
{\bf Abstract.}
In this paper, we give several simple methods
for drawing a whole rational surface (without base points)
as several B\'ezier patches.
The first two methods apply to surfaces specified
by  triangular control nets and partition the real projective plane
$\rprospac{2}$ into four and six triangles respectively. 
The third method applies to surfaces specified
by rectangular control nets and partitions the torus
$\rprospac{1}\times\rprospac{1}$ into four rectangular regions.
In all cases, the new control nets are obtained by sign flipping
and permutation of indices from the original control net.
The proofs that these formulae are correct involve
very little computations and instead exploit the geometry
of the parameter space ($\rprospac{2}$  or
$\rprospac{1}\times\rprospac{1}$).
We illustrate our method on some classical examples.
We also propose a new method for resolving base points using a
simple ``blowing up'' technique involving the computation of
``resolved'' control nets.
%\vfill\eject
%\tableofcontents
\vfill\eject
\section{Introduction}
\label{sec1}
In this paper, we consider the 
problem of drawing a whole rational surface.
For example, consider the sphere $F$ specified by the fractions
$$
x(u,v) = \frac{2u}{u^2 + v^2 + 1},\quad
y(u,v) = \frac{2v}{u^2 + v^2 + 1},\quad
z(u,v) = \frac{u^2 + v^2 - 1}{u^2 + v^2 + 1}.
$$
The  problem is that no matter how large the interval $[r, s]$
is, the trace $F([r, s]\times [r, s])$ of $F$ over $[r, s]\times [r, s]$ is not  the trace 
of the entire surface. In this particular example, we could take advantage
of symmetries, but in general, this may not be possible.
We could use any of the bijections from $\>]-1, 1[\>$ to
$\reals$ to reduce the parameter domain to the square
$[-1, 1]\times [-1, 1]$, but since these maps are at least quadratic,
this could triple the total degree of the surface, leading
to an impractical method. For example, using the map
$$t \mapsto \frac{t}{1 - t^2}$$
$$\hbox{the fraction}\quad\frac{1}{u^2 + uv}\quad\hbox{becomes}\quad
\frac{(1 - u^2)^2(1 - v^2)}{(1 - v^2)u^2 + (1 - u^2)uv}.$$

\medskip
Recomputing the control net after substitution
would also be quite expensive. Indeed, one of the reasons why the problem
is not trivial is that in most CAGD applications, the surface is
given in terms of control points rather than parametrically
(in terms of polynomials).

\medskip
Thus, the problem is to cope with the situation in which
$u$ or $v$ become infinite. But what do we mean exactly by that?
To deal with this situation rigorously, we can ``go projective'',
that is, homogenize the polynomials. However, this can be done
in two different ways. 
The first method is to homogenize
with respect to the total degree, replacing $u$ by $u/t$ and $v$ by $v/t$,
getting
$$
x = \frac{2ut}{u^2 + v^2 + t^2},\quad
y = \frac{2vt}{u^2 + v^2 + t^2},\quad
z = \frac{u^2 + v^2 - t^2}{u^2 + v^2 + t^2}.
$$
The parameter domain is now the real projective
plane $\rprospac{2}$. Points at infinity are the  points of homogeneous
coordinates $(u, v, 0)$ (i.e., when $t = 0$).
Observe that all these point at infinity
yield the north pole $(0, 0, 1)$ on the sphere.

\medskip
The second method 
is to homogenize separately  in $u$ and $v$, 
replacing $u$ by $u/t_1$  and  $v$ by $v/t_2$, getting
$$x = \frac{2ut_1t_2^2}{u^2t_2^2 + v^2t_1^2 + t_1^2t_2^2},\quad
y = \frac{2vt_1^2t_2}{u^2t_2^2 + v^2t_1^2 + t_1^2t_2^2},\quad
z = \frac{u^2t_2^2 + v^2t_1^2 - t_1^2t_2^2}{u^2t_2^2 + v^2t_1^2 + t_1^2t_2^2}.
$$
%$$
%\eqaligneno{
%x(u,v,t_1,t_2) &= \frac{2ut_1t_2^2}{u^2t_2^2 + v^2t_1^2 + t_1^2t_2^2},\cr
%y(u,v,t_1,t_2) &= \frac{2vt_1^2t_2}{u^2t_2^2 + v^2t_1^2 + t_1^2t_2^2},\cr
%z(u,v,t_1,t_2) &= 
%\frac{u^2t_2^2 + v^2t_1^2 - t_1^2t_2^2}{u^2t_2^2 + v^2t_1^2 + t_1^2t_2^2}.\cr
%}$$
This time, the parameter domain is the product space
$\rprospac{1}\times \rprospac{1}$, where $\rprospac{1}$ is the
real projective line. The domain 
$\rprospac{1}\times \rprospac{1}$ is homeomorphic to a torus,
and it is not homeomorphic to $\rprospac{2}$.
Observe that when $t_1 = t_2 = 0$,
all the numerators and the denominator vanish simultaneously.
We have what is called a {\it base point\/}.
This is annoying but not terribly surprising, since a sphere
is not of the same topological type as a torus.
It should be noted that there are also rational surfaces (such as the torus)
that do not have base points when treated as surfaces with domain
$\rprospac{1}\times \rprospac{1}$ but have base points when treated
as having domain $\rprospac{2}$ (see Section \ref{sec7}), and vice versa.

\medskip
In summary, there are two ways to deal with infinite
values of the parameters. 
We can homogenize with respect to the total degree $\mdeg$ 
(replacing $u$ by $u/t$ and $v$ by $v/t$). This leads to
rational surfaces specified by triangular control nets, as we will
see more precisely in the next section.
The other method is to homogenize
with respect to $u$ and the maximum degree $\pdeg$ in $u$ 
(replacing $u$ by $u/t_1$) and 
with respect to $v$ and the maximum degree $\qdeg$ in $v$
(replacing $v$ by $v/t_2$). 
This leads to
rational surfaces specified by rectangular control nets, as we will
see more precisely in the next section.

\medskip
The problem of drawing a rational surface
reduces to the problem of
partitioning the parameter domain into simple connected
regions $R_i$ such as triangles or rectangles, in such a way that
there is some prespecified region $R_0$ and some projectivities
such that every other region is the image of the 
region $R_0$ under one of the projectivities. Furthermore, if the
patch associated with the  region $R_0$ 
is given by a control net $\s{N}_{0}$,
we want the control net $\s{N}_{i}$ associated with the region
$R_i$ to be computable very quickly from  $\s{N}_{0}$.

\medskip
In the case of the real projective plane $\rprospac{2}$,
we can use the fact that  $\rprospac{2}$ is obtained as the
quotient of the sphere $S^2$ after identification of 
antipodal points. 
The real projective plane can be partitioned by
projecting  any polyhedron inscribed in the sphere $S^2$
on a plane. This way of dividing the real projective plane into
regions is discussed quite extensively 
in Hilbert and Cohn-Vossen \cite{Hilbert} (see Chapter III). 
As noted by Hilbert, it is better to use polyhedra with
central symmetry, so that the projective plane is covered
only once since  vertices come in pairs of antipodal points. 
In particular, we can use the four Platonic solids other
than the tetrahedron, but if we
want rectangular or triangular regions, only 
the cube, the octahedron, and the icosahedron  can be used.
Indeed, projection of the dodecahedron yields pentagonal
regions (see Hilbert and Cohn-Vossen \cite{Hilbert}, page 147-150).

\smallskip
If we  project the cube onto one of its faces from its center,
we get  three rectangular regions (see Section \ref{sec4}).
It is easy to find the projectivities that map the central
region onto the other two. 
Since we are dealing with the projective plane,
it is better to use triangular control nets to avoid base points, 
and it is necessary to split the central rectangle
into two triangles. Thus, the trace
of the rational surface is the union of six patches over
various triangles. It is  shown in Section \ref{sec4} how 
the control nets of the other four  patches are easily
(and cheaply) obtained from  the control nets of the two
central triangles.

\medskip
If we project the octahedron onto one of its faces from its center, 
we get four triangular regions (see Section \ref{sec5}).
This time, it is a little harder to write down the projectivities
that map the central triangle $rst$ to the other three triangles $R, S, T$.
However it is not necessary to find explicit formulae for these
projectivities, and using a geometric argument,
we can find very simple formulae to compute the
control nets associated with the other three triangles from the control net 
associated with the central triangle, as shown in Section \ref{sec5}.

\medskip
Projecting the icosahedron onto one of its faces from its center
yields ten triangular regions, but we haven't
found formulae to compute control nets of the other regions
from the central triangle. We leave the discovery
of such formulae as an open and possibly challenging problem. 

\medskip
Let us now consider the problem of partitioning
$\rprospac{1}\times \rprospac{1}$ 
into simple regions.
Since the projective line $\rprospac{1}$ is topologically a circle,
a very simple method is to inscribe a rectangle
(or a square) in the circle and then
project it. One way to do so leads to a partition of
$\rprospac{1}$ into $[-1, 1]$ and $\rprospac{1} - ]-1, 1[$.
The corresponding projectivity 
is $t \mapsto \frac{1}{t}$. 
Other projections lead to
a partition of $\rprospac{1}$ into $[r, s]$ and 
$\rprospac{1} - ]r, s[$ for any affine frame $(r, s)$.
In all cases, the torus is split into four rectangular regions,
and there are very simple formulae for computing
the control nets of the other three rectangular nets
from the control net associated with the patch over
$[-1, 1]\times [-1, 1]$ (or more generally, $[r_1, s_1]\times [r_2, s_2]$),
as explained in Section \ref{sec6}.

\medskip
It should be stressed that it is not necessary
to compute explicitly the various projectivities,
and that in each case, a simple geometric argument yields
the desired formulae for the new control nets.
Other methods for drawing rational surfaces were also investigated
by Bajaj and Royappa \cite{Bajaj94a,Bajaj95a} and 
DeRose \cite{deRose91} and  will be discussed in Section \ref{sec5}
and Section \ref{sec6}.

\medskip
There is a problem
with our methods when all the numerators and the denominator 
vanish simultaneously. In this case,
we have what is called a {\it base point\/}.
In Section \ref{sec7}, we give a new method for resolving base points
(in the case of a rational surface
specified by a triangular control net), using a
simple ``blowing-up'' technique based on an idea of Warren \cite{Warren92}.
What is new is that we give formulae for computing ``resolved'' control nets.

\medskip
It turns out that to give rigorous proofs of our formulae, it is necessary to
view rational surfaces as surfaces defined in a suitable projective
space in terms of multiprojective maps. We will summarize how to do this 
in Section \ref{sec3}. 
The proofs that our formulae are correct involve
very little computations and instead exploit the geometry
of the parameter space ($\rprospac{2}$  or
$\rprospac{1}\times\rprospac{1}$).
For the sake of brevity, we do not review how
polynomial surfaces are defined in terms of control points.
The deep reason why polynomial triangular surface patches can be
effectively handled in terms of control points is that
multivariate polynomials arise from multiaffine symmetric maps
(see Ramshaw \cite{Ramshaw87}, Farin \cite{Farin93,Farin95}, 
Hoschek and Lasser \cite{Hoschek}, 
or Gallier \cite{Gallbook}).

\section{Rational Surfaces and Control Points}
\label{sec3}
Denoting the affine plane $\affreal^2$ as $\s{P}$,
a rational surface $\mapdef{F}{\s{P}}{\reals^n}$ of degree
$\mdeg$ is specified by some fractions
$$
x_1(u,v) = \frac{F_1(u,v)}{F_{n+1}(u,v)},\quad\ldots\quad,
x_n(u,v) = \frac{F_n(u,v)}{F_{n+1}(u,v)},
$$
where $F_1(u,v), \ldots, F_{n+1}(u,v)$ are polynomials of total degree $\leq \mdeg$.
In order to handle rational surfaces in terms of control points,
it turns out that it is  necessary to view rational surfaces
as surfaces in some projective space. Roughly, this means that we
have to homogenize the polynomials $F_1(u,v), \ldots, F_{n+1}(u,v)$.
However, the polar forms of
homogeneous polynomials are multilinear, and thus we must deal
with multilinear maps rather than multiaffine maps.
Fortunately, there is a construction to embed an affine space into
a vector space, in such a way that multiaffine maps
extend uniquely to multilinear maps. This construction is described
in Berger \cite{Berger90} and is at the heart of the
presentation of rational surfaces in 
Fiorot and Jeannin \cite{Fiorot89,Fiorot92}.
However, Fiorot and Jeannin do not use polar forms. We have adapted
their approach in the framework of polar forms in Gallier \cite{Gallbook2}.
In this paper, we simply review the  facts needed to understand the proof of
our theorems. 
Given a vector space $E$, we denote the projective
space induced by $E$ as $\projs{E}$ (see  Berger \cite{Berger90} or 
Gallier \cite{Gallbook2}).
Given an affine space $E$ with associated vector space
$\vectorr{E}$, a vector space $\hli{E}$
can be constructed, such that $E$ is embedded as an affine
hyperplane in  $\hli{E}$ via an affine map
$\mapdef{j}{E}{\hli{E}}$, 
and $\vectorr{E}$ as a hyperplane. 

%
%\setbox\samplea\hbox{$(E, \vectorr{E})$}
%\setbox\sampleb\hbox{$\hli{E}$}
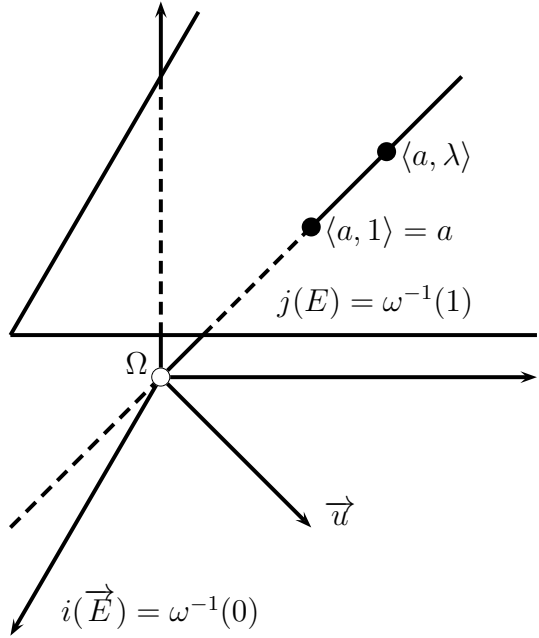
\begin{figure}[H]
%\begin{example}
  \begin{center}
    \begin{pspicture}(-3,-3.5)(5,4.8)
    \psline[linewidth=1.5pt]{->}(0,0)(-2,-3.44)
    \psline[linewidth=1.5pt]{->}(0,0)(5,0)
    \psline[linewidth=1.5pt](0,0)(0,0.56)
    \psline[linewidth=1.5pt,linestyle=dashed](0,0.56)(0,4)
    \psline[linewidth=1.5pt]{->}(0,4)(0,5)
    \psline[linewidth=1.5pt](-2,0.56)(5,0.56)
    \psline[linewidth=1.5pt](-2,0.56)(0.5,4.86)
    \psline[linewidth=1.5pt,linestyle=dashed](-2,-2)(0,0)
    \psline[linewidth=1.5pt](0,0)(0.56,0.56)
    \psline[linewidth=1.5pt,linestyle=dashed](0.56,0.56)(2,2)
    \psline[linewidth=1.5pt](2,2)(4,4)
    \psline[linewidth=1.5pt]{->}(0,0)(2,-2)
    \psdots[dotstyle=o,dotscale=2](0,0)
    \psdots[dotstyle=*,dotscale=2](3,3)
    \psdots[dotstyle=*,dotscale=2](2,2)
    \uput[150](0,0){$\Omega$}
    \uput[-10](2,2){$\pairt{a}{1} = a$}
    \uput[-10](3,3){$\pairt{a}{\lambda}$}
    \uput[-10](-1.5,-3){$i(\vectorr{E}) = \omega^{-1}(0)$}
    \uput[60](2,0.6){$j(E) = \omega^{-1}(1)$}
    \uput[30](2,-2){$\vectorr{u}$}
    \end{pspicture}
  \end{center}
  \caption{Embedding an affine space 
into a vector space}
%\end{example}
\end{figure}

Both hyperplanes are defined by some linear form
$\mapdef{\omega}{\hli{E}}{\reals}$.
The previous diagram illustrates the embedding of the affine space $E$
into the vector space $\hli{E}$:
Roughly, the vector space $\hli{E}$ has the property that
for any vector space $\vectorr{F}$ and any affine
map  $\mapdef{f}{E}{\vectorr{F}}$, there is a unique
linear map  $\mapdef{\hli{f}}{\hli{E}}{\vectorr{F}}$
extending $\mapdef{f}{E}{\vectorr{F}}$. As a consequence,
given two affine spaces $E$ and $F$,
every affine map $\mapdef{f}{E}{F}$
extends uniquely to a linear map 
$\mapdef{\hli{f}}{\hli{E}}{\hli{F}}$.

\medskip
A pair $\pairt{a}{\lambda}$ where $a\in E$ and $\lambda\not=0$ is called
a {\it weighted point\/}. Vectors in $\hli{E}$ of the form
$\pairt{a}{1}$ are identified with points $a$ in $E$.
It is easily shown that for every
$a\in E$, we have $\hli{E} = \vectorr{E}\oplus \reals \ptb{a}$. 

\medskip
We have the following important result whose proof can be found
in Gallier  \cite{Gallbook2}, or inferred from Ramshaw \cite{Ramshaw87}.

\begin{lemma}
\label{homogaff2}
Given any two affine spaces $E$ and $F$ and a multiaffine map
$\mapdef{f}{E^m}{F}$, there is a unique multilinear map 
$\mapdef{\hli{f}}{(\hli{E})^m}{\hli{F}}$ extending $f$
as in the diagram below: 
%
%\[
%\matrice{
%\begin{matrix}
%E^{\mdeg}        & \maprightu{f}          & F     \\
%\mapdownl{j\times\cdots\times j}  &          & \mapdownr{j} \\
%(\hli{E})^{\mdeg}     & \maprightd{\hli{f}}      & \hli{F}         
%\end{matrix}
%}
%\]
\[
\xymatrix{
E^{m}    \ar[r]^{f} \ar[d]_-{j\times\cdots\times j}   &     \>  F 
\ar[d]^-{j}    \\
(\hli{E})^{\mdeg}    \ar[r]_-{\hli{f}}      & \> \hli{F}    .     
}
\]
\end{lemma}

\medskip
Given an affine space $E$,
the projective space $\projs{\hli{E}}$ induced by $\hli{E}$
is denoted as $\pcompl{E}$, and it is called the
{\it projective completion of $E$\/}. Observe that
$\pcompl{\reals} = \rprospac{1}$ and $\pcompl{\s{P}} = \rprospac{2}$.

\medskip
The upshot of the above considerations is that
a rational surface can be defined in terms of multilinear maps.
Let $\affs$ be some ambiant affine space in which
our surfaces live, in most cases $\reals^3$.
If we first homogenize the polynomials $F_i(u, v)$
with respect to the total degree $\mdeg$ (replacing $u$ by $u/z$ and $v$ by $v/z$),
we can view a rational surface as a map
$$\mapdef{F}{\pcompl{\s{P}}}{\pcompl{\affs}}$$
(where $\s{P} = \affreal^2$ is the affine plane)
induced by some  symmetric multilinear map
$$\mapdef{f}{(\hli{\s{P}})^{\mdeg}}{\hli{\affs}}$$
such that
$$F([u, v, z]) = \projs{f}(\underbrace{(u, v, z),\ldots,(u, v, z)}_{\mdeg}),$$
for all homogeneous coordinates $(u, v, z)\in\reals^3$.
We call such  surfaces {\it rational total degree surfaces\/}, or
{\it triangular rational surfaces\/}.
Furthermore, for any affine frame $\Delta rst$,
the {\it triangular control net\/} 
$\s{N}= (\theta_{i,\,j,\,k})_{(i,j,k)\in\Delta_{\mdeg}}$ 
(in $\hli{\affs}$) w.r.t. $\Delta rst$  defining the triangular surface
$F$ is given by the formulae
$$\theta_{i,\,j,\,k} = f(\underbrace{r, \ldots, r}_{i},
                         \underbrace{s, \ldots, s}_{j},
                         \underbrace{t, \ldots, t}_{k}),\quad
\hbox{where}\> i + j + k = \mdeg.$$

On the other hand, if we first homogenize the polynomials $F_i(u, v)$
with respect to $u$ and the maximum degree $\pdeg$ in $u$ 
(replacing $u$ by $u/t_1$) and second
with respect to $v$ and the maximum degree $\qdeg$ in $v$
(replacing $v$ by $v/t_2$),
we can view a rational surface as a map
$$\mapdef{F}{\pcompl{\affreal}\times\pcompl{\affreal}}{\pcompl{\affs}}$$
induced by some  multilinear map
$$\mapdef{f}{(\hli{\affreal})^{\pdeg}\times (\hli{\affreal})^{\qdeg}}{\hli{\affs}}$$
which is symmetric in its first $\pdeg$ arguments and in its last $\qdeg$ arguments,
and such that
$$F([u, t_1], [v, t_2]) = 
\projs{f}(\underbrace{(u, t_1),\ldots,(u, t_1)}_{\pdeg},
\underbrace{(v, t_2),\ldots,(v, t_2)}_{\qdeg}),$$
for all homogeneous coordinates $(u, t_1), (v, t_2)\in\reals^2$.
We call such  surfaces {\it rational  surfaces of bidegree $\pairt{\pdeg}{\qdeg}$\/}, or
{\it rectangular rational surfaces\/}.
Furthermore, given any two affine frames $(r_1, s_1)$ and $(r_2, s_2)$
for the affine line $\reals$, 
the {\it rectangular control net\/}
$\s{N} = (\theta_{i,\, j})_{0\leq i\leq \pdeg,\, 0\leq j\leq \qdeg}$ 
(in $\hli{\affs}$)  w.r.t. $(r_1, s_1)$ and $(r_2, s_2)$  
defining the rectangular surface
$F$ is given by the formulae
$$\theta_{i,\,j} = f(\underbrace{r_1, \ldots, r_1}_{\pdeg - i},
                         \underbrace{s_1, \ldots, s_1}_{i},
                     \underbrace{r_2, \ldots, r_2}_{\qdeg - j},
                         \underbrace{s_2, \ldots, s_2}_{j}).
$$
A {\it base point\/} of a rational surface $F$
specified by a multilinear map $f$ is any point
$a\in\reals^3$ such that
$f(\underbrace{a, \ldots, a}_{\mdeg}) = \vectorr{0}$,
or any point $(u, v)\in\reals^2$ such that
$f(\underbrace{u, \ldots, u}_{\pdeg},
\underbrace{v, \ldots, v}_{\qdeg}) = \vectorr{0}$.

\section{Splitting Triangular Rational Surfaces Into Six Triangular Patches}
\label{sec4}
As we explained in Section \ref{sec1}, 
if we project a cube onto one of its faces from its center,
we obtain a partition of the projective plane $\rprospac{2}$
into three rectangular regions, in such a way that there exist simple projectivities
$\varphi$ and $\psi$ between the square $[-1, 1]\times [-1, 1]$ and
the other two regions.

\begin{figure}[H]
\centerline{
\epsfig{file=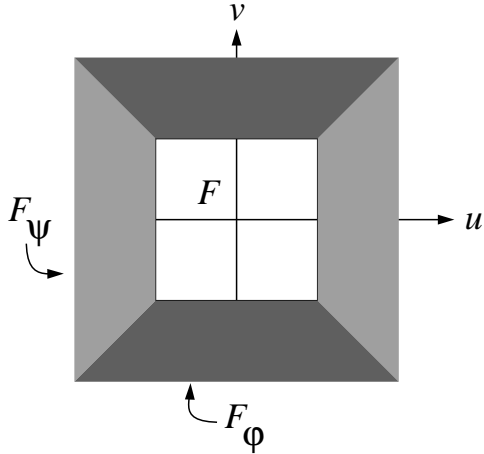,width=2.5truein}}
\caption{Dividing the projective plane into $3$ rectangular regions}
\end{figure}

The projectivity $\varphi$ is
induced by the linear isomorphism of $\reals^3$ given by
$$(u, v, w) \mapsto (v, w, u).$$
Choosing the line at infinity $w = 0$,
the restriction of this map to 
the affine  plane $\s{P}$ (corresponding to $w=1$) is the map
$(u, v, 1) \mapsto (v, 1, u)$.
This is the map from $\s{P}$ to $\pcompl{\s{P}}$ such that
$(u, v) \mapsto (v/u, 1/u)$
if $u\not= 0$, and $(0, v) \mapsto (v, 1, 0)$ when $u = 0$.
The projectivity  $\psi$ is
induced by the linear isomorphism of $\reals^3$ given by
$$(u, v, w) \mapsto (w, u, v).$$
Choosing the line at infinity $w = 0$,
the restriction of this map to 
the affine plane $\s{P}$ (corresponding to $w=1$) is the map
$(u, v, 1) \mapsto (1, u, v)$.
This is the map from $\s{P}$ to $\pcompl{\s{P}}$ such that
$(u, v) \mapsto (1/v, u/v)$
if $v\not= 0$, and $(u, 0) \mapsto (1, u, 0)$ when $v = 0$.

\medskip
Actually, it turns out that the method of this section
holds for any region defined by a  nondegenerate quadrilateral
$(a, b, c, d)$, 
i.e. when $(a, b, c, d)$ is a projective frame.
However, the details are a bit messy, and for simplicity,
we restrict out attention to 
a rectangular region $[r_1, s_1]\times [r_2, s_2]$.
Since we are dealing with triangular surfaces,
it will be necessary to split the rectangle $[r_1, s_1]\times [r_2, s_2]$
into two triangles, and thus, we will obtain the trace
of a rational surface as the union of $6$ patches over
various triangles in the rectangle  $[r_1, s_1]\times [r_2, s_2]$.
Letting $a, b, c, d$ be the vertices of the rectangle $[r_1, s_1]\times [r_2, s_2]$
defined such that
$$
a = (s_1, s_2),\quad
b = (r_1, s_2),\quad
c = (r_1, r_2),\quad
d = (s_1, r_2),
$$
as shown below
 
\begin{figure}[H]
%\begin{example}
  \begin{center}
    \begin{pspicture}(-3,-3.2)(3,3)
    \psline[linewidth=1.5pt](-2,-2)(2,2)
    \psline[linewidth=1.5pt](-2,2)(2,-2)
    \psline[linewidth=1.5pt](-2,-2)(-2,2)
    \psline[linewidth=1.5pt](2,-2)(2,2)
    \psline[linewidth=1.5pt](-2,-2)(2,-2)
    \psline[linewidth=1.5pt](-2,2)(2,2)
    \psline[linewidth=1pt](-3,0)(3,0)
    \psline[linewidth=1pt](0,-3)(0,3)
    \psdots[dotstyle=o,dotscale=1.5](0,0)
    \uput[45](2,2){$a$}
    \uput[135](-2,2){$b$}
    \uput[-135](-2,-2){$c$}
    \uput[-45](2,-2){$d$}
    \uput[0](3,0){$u$}
    \uput[90](0,3){$v$}
    \end{pspicture}
  \end{center}
  \caption{Some affine frames associated with the rectangle 
 $(a, b, c, d)$}
%\end{example}
\end{figure}
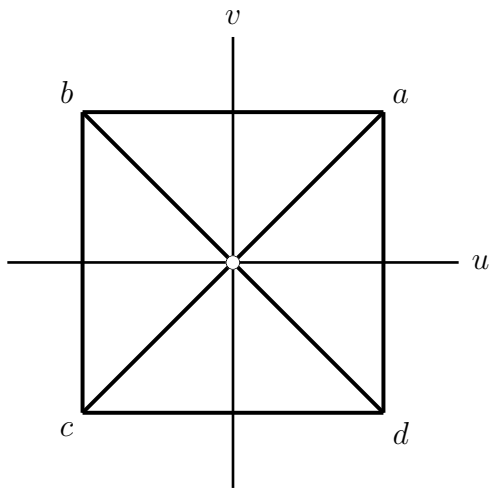

\noindent
we will consider the following affine frames
$$\eqaligneno{
\Delta bca &= ((r_1, s_2),\> (r_1, r_2),\> (s_1, s_2)),\cr
\Delta dac &= ((s_1, r_2),\> (s_1, s_2),\> (r_1, r_2)),\cr
\Delta bad &= ((r_1, s_2),\> (s_1, s_2),\> (s_1, r_2)).\cr
}$$

In particular, a rectangular surface patch defined
over the rectangle $[r_1, s_1]\times [r_2, s_2]$ will be treated as
the union of two triangular surface patches defined over the 
triangles $\Delta bca$ and $\Delta dac$.
It is somewhat unfortunate that a control net over 
the third frame $\Delta bad$ needs to be computed, but 
that is what the proof of lemma \ref{rendlem2} shows.
In any case, such a control net can be computed very cheaply
from a control net over $\Delta bca$ (or $\Delta dac$).

\medskip
There is simple geometric explanation of
the partitioning method in terms of the 
usual model of the real projective plane $\pcompl{\s{P}} = \rprospac{2}$
in $\reals^3$. Recall that in this model, the real projective plane
$\rprospac{2}$ consists of the points in
the plane $z = 1$ corresponding to the lines through the origin
not in the plane $z = 0$, and of the points at infinity corresponding
to the lines through the origin in the plane $z = 0$.
We view the vertices of the rectangle $(a, b, c, d)$ defined above
as points in the plane $z = 1$, in which case their
coordinates are $(s_1, s_2, 1)$, $(r_1, s_2, 1)$, $(r_1, r_2, 1)$, and
$(s_1, r_2, 1)$.  Then, we have the parallelepiped
$(a, b, c, d, -a, -b, -c, -d)$.
There is a unique projectivity $\projs{\varphi}$ such that
$$
\projs{\varphi}(a) = a,\quad \projs{\varphi}(b) = c,\quad 
\projs{\varphi}(c) = d,\quad \projs{\varphi}(d) = b.$$
For instance, it is induced by the 
unique linear  map $\varphi$ such that
$$
\varphi(a) = a,\quad
\varphi(b) = -c,\quad
\varphi(c) = -d.
$$
Since $d = -b + a + c$, we get
$$\varphi(d) = -\varphi(b) + \varphi(a) + \varphi(c) =
c + a - d = (r_1, s_2, 1) = b.$$
The linear map $\varphi$ transforms the top face $(a, b, c, d)$
of the parallelepiped to the back face $(a, -c, -d, b)$.
When a line $L$ through the origin and passing through a point of the face
$(a, -c, -d, b)$ varies, the intersection of $L$ with 
the plane $z = 1$ varies in $\varphi([r_1, s_1]\times [r_2, s_2])$.
We can define a rhombus
$(a, e, f, g, -a, -e, -f, -g)$  inscribed in the
sphere of center $O = (0, 0, 0)$ and of
radius $R = \sqrt{s_1^2 + s_2^2 + 1}$ passing through $a$,
as follows:
the points $e, f, g$ are on the upper half-sphere and
they are determined by the intersection of the
lines $(O, b)$, $(O, c)$ and $(O, d)$ with the sphere.
Then, it is obvious that under the central projection
of center $O$ onto the plane $z = 1$,
the top face $(a, e, f, g)$ of the rhombus projects onto the face $(a, b, c, d)$
of the parallelepiped, and that the projection of
the rhombus onto the plane $z = 1$ yields the desired
partitioning of $\rprospac{2}$.

\begin{figure}[H]
\centerline{
\psfig{file=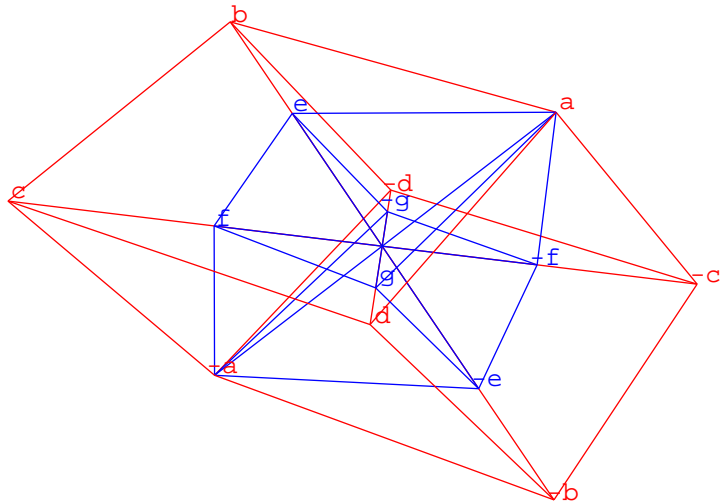,width=4truein}}
\caption{Parallelepiped and rhombus associated with $(a, b, c, d)$}
\label{rp2b}
\end{figure}

\medskip
Similarly, there is a unique projectivity $\projs{\psi}$ such that
$$
\projs{\psi}(a) = a,\quad \projs{\psi}(b) = d,\quad 
\projs{\psi}(c) = b,\quad \projs{\psi}(d) = c.$$
It is induced by the unique linear map $\psi$ such that
$$
\psi(a) = a,\quad
\psi(b) = d,\quad
\psi(c) = -b.
$$
Since $d = -b + a + c$, we get
$$\psi(d) = -\psi(b) + \psi(a) + \psi(c) =
-d + a - b = (-r_1, -r_2, -1) = -c.$$
The linear map $\psi$ transforms the top face $(a, b, c, d)$
of  the parallelepiped  to the right face $(a, d, -b, -c)$.
When a line $L$ through the origin and passing through a point of the face
$(a, d, -b, -c)$ varies, the intersection of $L$ with 
the plane $z = 1$ varies in $\psi([r_1, s_1]\times [r_2, s_2])$.
Again, it is obvious that under the central projection
of center $O$ onto the plane $z = 1$,
the top face $(a, e, f, g)$ of the rhombus projects onto the face $(a, b, c, d)$
of the parallelepiped, and that the projection of
the rhombus onto the plane $z = 1$ yields the desired
partitioning of $\rprospac{2}$.
Figure \ref{rp2b} shows the parallelepiped 
$(a, b, c, d, -a, -b, -c, -d)$
and the rhombus $(a, e, f, g, -a, -e, -f, -g)$.

\medskip
We will now use the maps $\varphi$ and $\psi$ to
show how the trace of a rational surface $F$ can be obtained
as the union of the traces of  three rational surfaces 
over the rectangle $[r_1, s_1]\times [r_2, s_2]$.%
\footnote{
While reading Appell's Treatise of Rational Mechanics,
we stumbled on the fact that  the change of variable
$(u, v) \mapsto (1/v, u/v)$
was used by Appell in his solution to a problem of Bertrand
(see \cite{Appell1}, Tome I, Part III, Chapter XI, page 422-423).
Appell explains that he found this ``homographic transformation''
in 1889. The problem of Bertrand is to find all central force laws
depending only on the position of a moving particle, so that the
trajectory of the particle is a conic for every choice of
initial conditions.}
The first of these surfaces is $F$ itself, and the
two other rational surfaces $F_{\varphi}$ and $F_{\psi}$
are easily obtained from $F$. However, depending on the 
multilinear map $f$ defining $F$, the surface $F$
(and thus, $F_{\varphi}$ and $F_{\psi}$) 
may have {\it base points\/}, that is,
we may have
$$f(\underbrace{(u, v, z),\ldots,(u, v, z)}_{\mdeg}) = \vectorr{0}$$
for some $(u, v, z)\not= (0, 0, 0)$.
We will show how to deal with this
situation later on.

\medskip
In order to render the trace of $F$, we will use the fact that
it is the union of the six traces $F(\Delta bca)$,
$F(\Delta dac)$, $F(\varphi(\Delta bca))$, $F(\varphi(\Delta dac))$,
$F(\psi(\Delta bca))$, and $F(\psi(\Delta dac))$.
Furthermore, the last four traces are also obtained
as traces of  $F_{\varphi}$ and $F_{\psi}$ over some
appropriate choice of affine frames among
$\Delta bca$, $\Delta dac$, and $\Delta bad$.

\medskip
We now show how $F_{\varphi}$ and $F_{\psi}$ are defined,
and how their control points can be computed very simply
from the control points of $F$ (computed with respect to the 
affine frames $\Delta bca$, $\Delta dac$, and $\Delta bad$).
We will assume
that the homogenization $\hli{\s{P}}$ of the affine plane
$\s{P}$ is identified with the direct sum
$\reals^2 \oplus \reals O$, where $O = (0, 0)$.
Then, every element of
$\hli{\s{P}}$ is of the form $(u, v, z)\in\reals^3$.

\begin{defin}
\label{fphispisdef}
{\em
Given an affine space $\affs$ of dimension $\geq 3$,
for every rational surface
$\mapdef{F}{\pcompl{\s{P}}}{\pcompl{\affs}}$
of degree $\mdeg$ specified by 
some symmetric multilinear map 
$\mapdef{f}{(\hli{\s{P}})^{\mdeg}}{\hli{\affs}}$,
the symmetric multilinear maps
$\mapdef{f_{\varphi}}{(\hli{\s{P}})^{\mdeg}}{\hli{\affs}}$
and $\mapdef{f_{\psi}}{(\hli{\s{P}})^{\mdeg}}{\hli{\affs}}$
are defined such that
$$\eqaligneno{
f_{\varphi}((u_{1}, v_{1}, w_{1}),
\ldots,(u_{\mdeg}, v_{\mdeg}, w_{\mdeg})) &= 
f(\varphi(u_{1},\> v_{1},\> w_{1}),
\ldots, \varphi(u_{\mdeg},\> v_{\mdeg},\> w_{\mdeg})),\cr
f_{\psi}((u_{1}, v_{1}, w_{1}),
\ldots,(u_{\mdeg}, v_{\mdeg}, w_{\mdeg})) &=
f(\psi(u_{1},\> v_{1},\> w_{1}), \ldots,
\psi(u_{\mdeg},\> v_{\mdeg},\> w_{\mdeg})).\cr
}$$
Let $\mapdef{F_{\varphi}}{\pcompl{\s{P}}}{\pcompl{\affs}}$
be the rational surface specified by
$\mapdef{f_{\varphi}}{(\hli{\s{P}})^{\mdeg}}{\hli{\affs}}$, and let
$\mapdef{F_{\psi}}{\pcompl{\s{P}}}{\pcompl{\affs}}$
be the rational surface specified by
$\mapdef{f_{\psi}}{(\hli{\s{P}})^{\mdeg}}{\hli{\affs}}$.
}
\end{defin}

Observe that the base points of $F_{\varphi}$, if any, 
have coordinates $(u, v, w)\not=(0, 0, 0)$  such that
$$f(\varphi(u,\> v,\> w), \ldots, \varphi(u,\> v,\> w)) = \vectorr{0},$$
and that the base points of $F_{\psi}$, if any, 
have coordinates $(u, v, w)\not= (0, 0, 0)$ such that
$$f(\psi(u,\> v, \> w), \ldots, \psi(u,\> v,\> w)) = \vectorr{0}.$$

\begin{lemma}
\label{rendlem2}
Given an affine space $\affs$ of dimension $\geq 3$,
for every rational surface
$\mapdef{F}{\pcompl{\s{P}}}{\pcompl{\affs}}$
of degree $\mdeg$ specified by 
some symmetric multilinear map 
$\mapdef{f}{(\hli{\s{P}})^{\mdeg}}{\hli{\affs}}$,
if $f_{\varphi}$ and $f_{\psi}$
are the symmetric multilinear maps of definition 
\ref{fphispisdef}, except for  base points, $F$, $F_{\varphi}$ and $F_{\psi}$ 
have the same trace.
The trace of $F_{\varphi}$  over $\Delta bca$ 
is the trace of $F$ over $\varphi(\Delta bca)$,
the trace of $F_{\varphi}$ over $\Delta dac$ is the trace 
of $F$ over $\varphi(\Delta dac)$,
the trace of $F_{\psi}$  over $\Delta bca$  is the trace 
of $F$  over $\psi(\Delta bca)$,
and the trace of $F_{\psi}$ over $\Delta dac$  is the trace 
of $F$  over $\psi(\Delta dac)$.
Furthermore, if the control nets 
(in $\hli{\affs}$) of the surface $F$
w.r.t. the affine frames $\Delta bca$, $\Delta dac$, and $\Delta bad$,
are respectively 
$$\eqaligneno{
\alpha &= (\alpha_{i,\,j,\,k})_{(i,j,k)\in\Delta_{\mdeg}},\cr
\beta &= (\beta_{i,\,j,\,k})_{(i,j,k)\in\Delta_{\mdeg}},\cr
\gamma &= (\gamma_{i,\,j,\,k})_{(i,j,k)\in\Delta_{\mdeg}},\cr
}$$
the control nets $\theta^{1}$ and $\theta^{2}$ (in $\hli{\affs}$)
of the surface $F_{\varphi}$ 
w.r.t. the affine frames $\Delta bca$ and $\Delta dac$,
and the control nets $\rho^{1}$ and $\rho^{2}$ (in $\hli{\affs}$)
of the surface $F_{\psi}$
w.r.t. the affine frame $\Delta bca$ and $\Delta dac$,
are given by the equations
$$\eqaligneno{
\theta_{i,\, j,\, k}^{1} &= (-1)^{i + j}\> \beta_{j,\, k,\, i},\cr
\theta_{i,\, j,\, k}^{2} &= (-1)^{k}\> \gamma_{i,\, j,\, k},\cr
\rho_{i,\, j,\, k}^{1} &= (-1)^{j}\> \gamma_{j,\, k,\, i},\cr
\rho_{i,\, j,\, k}^{2} &= (-1)^{i + k}\> \alpha_{k,\, i,\, j}.\cr
}$$
\end{lemma}

%\medskip
\proof 
%$$f_{\varphi}((u_{1}, v_{1}, w_{1}),
%\ldots,(u_{\mdeg}, v_{\mdeg}, w_{\mdeg})) = 
%u_{1}\cdots u_{\mdeg}
%f\left(\varphi\left(1,\> \frac{v_{1}}{u_{1}},
%\> \frac{w_{1}}{u_{1}}\right),
%\ldots,\varphi\left(1,\> \frac{v_{\mdeg}}{u_{\mdeg}},
%\> \frac{w_{\mdeg}}{u_{\mdeg}}\right)\right),$$
%and thus, we get
%$$\projs{f_{\varphi}}([u_{1}, v_{1}, w_{1}],
%\ldots,[u_{\mdeg}, v_{\mdeg}, w_{\mdeg}]) =
%\projs{f}\left(\left[\varphi\left(1,\> \frac{v_{1}}{u_{1}},\, 
%\frac{w_{1}}{u_{1}}\right)\right], \ldots,
%\left[\varphi\left(1,\> \frac{v_{\mdeg}}{u_{\mdeg}},\, 
%\frac{w_{\mdeg}}{u_{\mdeg}}\right)\right]\right).$$ 
We have
$$f_{\varphi}((u_{1}, v_{1}, w_{1}),
\ldots,(u_{\mdeg}, v_{\mdeg}, w_{\mdeg})) = 
f(\varphi(u_{1}, v_{1}, w_{1}),
\ldots,\varphi(u_{\mdeg}, v_{\mdeg}, w_{\mdeg})),$$
and thus
$$\projs{f_{\varphi}}([u_{1}, v_{1}, w_{1}],
\ldots,[u_{\mdeg}, v_{\mdeg}, w_{\mdeg}]) = 
\projs{f}([\varphi(u_{1}, v_{1}, w_{1})],
\ldots,[\varphi(u_{\mdeg}, v_{\mdeg}, w_{\mdeg})]).$$
In view of the properties of $\varphi$,
it is clear that $F$ and $F_{\varphi}$ have the same
trace (except for base points), 
and that the trace of $F_{\varphi}$ over 
$\Delta bca$  is the trace of $F$ 
over $\varphi(\Delta bca)$,
and  the trace of $F_{\varphi}$  over $\Delta dac$ is the trace 
of $F$  over $\varphi(\Delta dac)$.
A similar argument applies to $F$ and $F_{\psi}$.
The formulae for computing the control points of
$F_{\varphi}$ w.r.t. the triangle $\Delta bca$ are obtained
by computing 
$$f_{\varphi}(\underbrace{b, \ldots, b}_{i},
\underbrace{c, \ldots, c}_{j},
\underbrace{a, \ldots, a}_{k}).$$
Since
$$f_{\varphi}((u_{1}, v_{1}, w_{1}),
\ldots,(u_{\mdeg}, v_{\mdeg}, w_{\mdeg})) = 
f(\varphi(u_{1},\> v_{1},\> w_{1}),
\ldots, \varphi(u_{\mdeg},\> v_{\mdeg},\> w_{\mdeg})),$$
$\varphi(b) = -c$, $\varphi(c) = -d$, and $\varphi(a) = a$,
we have
$$f_{\varphi}(\underbrace{b, \ldots, b}_{i},
\underbrace{c, \ldots, c}_{j},
\underbrace{a, \ldots, a}_{k})
= f(\underbrace{-c, \ldots, -c}_{i},
\underbrace{-d, \ldots, -d}_{j},
\underbrace{a, \ldots, a}_{k}),$$ 
that is
$$f_{\varphi}(\underbrace{b, \ldots, b}_{i},
\underbrace{c, \ldots, c}_{j},
\underbrace{a, \ldots, a}_{k}) =
(-1)^{i + j}f(\underbrace{c, \ldots, c}_{i},
\underbrace{d, \ldots, d}_{j},
\underbrace{a, \ldots, a}_{k}),$$
and since the control points  $\beta_{i,\, j,\, k}$ are computed w.r.t.
the triangle $\Delta dac$, we get 
$$\theta_{i,\, j,\, k}^{1} = (-1)^{i + j}\> \beta_{j,\, k,\, i}.$$
The formulae for computing the control points of
$F_{\varphi}$ w.r.t. the triangle $\Delta dac$ are obtained
by computing 
$$f_{\varphi}(\underbrace{d, \ldots, d}_{i},
\underbrace{a, \ldots, a}_{j},
\underbrace{c, \ldots, c}_{k}).$$
Since
$$f_{\varphi}((u_{1}, v_{1}, w_{1}),
\ldots,(u_{\mdeg}, v_{\mdeg}, w_{\mdeg})) = 
f(\varphi(u_{1},\> v_{1},\> w_{1}),
\ldots, \varphi(u_{\mdeg},\> v_{\mdeg},\> w_{\mdeg})),$$
$\varphi(d) = b$, $\varphi(c) = -d$, and $\varphi(a) = a$,
we have
$$f_{\varphi}(\underbrace{d, \ldots, d}_{i},
\underbrace{a, \ldots, a}_{j},
\underbrace{c, \ldots, c}_{k})
= f(\underbrace{b, \ldots, b}_{i},
\underbrace{a, \ldots, a}_{j},
\underbrace{-d, \ldots, -d}_{k}),$$ 
that is
$$f_{\varphi}(\underbrace{d, \ldots, d}_{i},
\underbrace{a, \ldots, a}_{j},
\underbrace{c, \ldots, c}_{k}) =
(-1)^{k}f(\underbrace{b, \ldots, b}_{i},
\underbrace{a, \ldots, a}_{j},
\underbrace{d, \ldots, d}_{k}),$$
and since the control points  $\gamma_{i,\, j,\, k}$ are computed w.r.t.
the triangle $\Delta bad$, we get 
$$\theta_{i,\, j,\, k}^{2} = (-1)^{k}\> \gamma_{i,\, j,\, k}.$$
The formulae for computing the control points of
$F_{\psi}$ w.r.t. the triangle $\Delta bca$ are obtained
by computing 
$$f_{\psi}(\underbrace{b, \ldots, b}_{i},
\underbrace{c, \ldots, c}_{j},
\underbrace{a, \ldots, a}_{k}).$$
Since
$$f_{\psi}((u_{1}, v_{1}, w_{1}),
\ldots,(u_{\mdeg}, v_{\mdeg}, w_{\mdeg})) = 
f(\psi(u_{1},\> v_{1},\> w_{1}),
\ldots, \psi(u_{\mdeg},\> v_{\mdeg},\> w_{\mdeg})),$$
$\psi(b) = d$, $\psi(c) = -b$, and $\psi(a) = a$,
we have
$$f_{\psi}(\underbrace{b, \ldots, b}_{i},
\underbrace{c, \ldots, c}_{j},
\underbrace{a, \ldots, a}_{k})
= f(\underbrace{d, \ldots, d}_{i},
\underbrace{-b, \ldots, -b}_{j},
\underbrace{a, \ldots, a}_{k}),$$ 
that is
$$f_{\psi}(\underbrace{b, \ldots, b}_{i},
\underbrace{c, \ldots, c}_{j},
\underbrace{a, \ldots, a}_{k}) =
(-1)^{j}f(\underbrace{d, \ldots, d}_{i},
\underbrace{b, \ldots, b}_{j},
\underbrace{a, \ldots, a}_{k}),$$
and since the control points  $\gamma_{i,\, j,\, k}$ are computed w.r.t.
the triangle $\Delta bad$, we get 
$$\rho_{i,\, j,\, k}^{1} = (-1)^{j}\> \gamma_{j,\, k,\, i}.$$
Finally, the formulae for computing the control points of
$F_{\psi}$ w.r.t. the triangle  $\Delta dac$ are obtained
by computing 
$$f_{\psi}(\underbrace{d, \ldots, d}_{i},
\underbrace{a, \ldots, a}_{j},
\underbrace{c, \ldots, c}_{k}).$$
Since
$$f_{\psi}((u_{1}, v_{1}, w_{1}),
\ldots,(u_{\mdeg}, v_{\mdeg}, w_{\mdeg})) = 
f(\psi(u_{1},\> v_{1},\> w_{1}),
\ldots, \psi(u_{\mdeg},\> v_{\mdeg},\> w_{\mdeg})),$$
$\psi(d) = -c$, $\psi(c) = -b$, and $\psi(a) = a$,
we have
$$f_{\psi}(\underbrace{d, \ldots, d}_{i},
\underbrace{a, \ldots, a}_{j},
\underbrace{c, \ldots, c}_{k})
= f(\underbrace{-c, \ldots, -c}_{i},
\underbrace{a, \ldots, a}_{j},
\underbrace{-b, \ldots, -b}_{k}),$$ 
that is
$$f_{\psi}(\underbrace{d, \ldots, d}_{i},
\underbrace{a, \ldots, a}_{j},
\underbrace{c, \ldots, c}_{k}) =
(-1)^{i + k}f(\underbrace{c, \ldots, c}_{i},
\underbrace{a, \ldots, a}_{j},
\underbrace{b, \ldots, b}_{k}),$$
and since the control points  $\alpha_{i,\, j,\, k}$ are computed w.r.t.
the triangle $\Delta bca$, we get 
$$\rho_{i,\, j,\, k}^{2} = (-1)^{i + k}\> \alpha_{k,\, i,\, j}.$$
$\square$

\medskip
The above calculations show that $\varphi$ and $\psi$
can be defined as above provided that
$d = -b + a + c$, or equivalently 
$b + d = a + c$, which means that $(a, b, c, d)$ is a parallelogram.
Actually, lemma \ref{rendlem2}  
also holds in the more general situation
where $(a, b, c, d)$ is a projective frame, i.e.
a quadrilateral whose vertices are in general position.
However, the definition of the linear maps $\varphi$
and $\psi$ is a little more messy. As before, we identify
$a, b, c, d$ with points in the plane $z = 1$,
and we let $a = (a_1, a_2, 1)$, $b = (b_1, b_2, 1)$,
$c = (c_1, c_2, 1)$, and $d = (d_1, d_2, 1)$.
To find a linear map $\varphi$ inducing the unique projectivity
$\projs{\varphi}$ such that
$$
\projs{\varphi}(a) = a,\quad \projs{\varphi}(b) = c,\quad 
\projs{\varphi}(c) = d,\quad \projs{\varphi}(d) = b,$$
we let $d = \lambda a + \mu b + \nu c$
and $b = \lambda' a + \mu' c + \nu' d$, where
$\lambda + \mu + \nu = 1$ and $\lambda' + \mu' + \nu' = 1$,
and $\varphi$ is the unique linear map such that
$$
\varphi(\lambda a) = \lambda' a,\quad
\varphi(\mu b) = \mu' c,\quad
\varphi(\nu c) = \nu' d.$$
Then,
$\varphi(d) = b$, as desired. The linear map $\psi$ 
can be defined in a similar way.
The proof still goes through since
the maps involved are multilinear, and thus not disturbed by scalar
multiples.

\medskip
Lemma \ref{rendlem2} shows that in order to render a rational surface,
provided that it does not have base points,
we just need to compute the control nets $\alpha, \beta, \gamma$
for the surface $F$ w.r.t. the affine frames $\Delta bca$, 
$\Delta dac$, and $\Delta bad$, since then,
the control nets $\theta^{1}$ and $\theta^{2}$ (in $\hli{\affs}$)
of the surface $F_{\varphi}$ 
w.r.t. the affine frames $\Delta bca$ and $\Delta dac$,
and the control nets $\rho^{1}$ and $\rho^{2}$ (in $\hli{\affs}$)
of the surface $F_{\psi}$
w.r.t. the affine frame $\Delta bca$ and $\Delta dac$,
are obtained at trivial cost.

%\medskip
\remark 
It should be noted that the surface patches
associated with the control nets $\alpha$, $\beta$,
$\theta^{1}$, $\theta^{2}$, $\rho^{1}$, and $\rho^{2}$,
may overlap in more than boundaries.
In fact, there are examples where $\alpha$ and $\beta$
determine the entire surface, and other examples in which
$\theta^{1}$, $\theta^{2}$, $\rho^{1}$, and $\rho^{2}$,
determine the entire surface.

\medskip
It is fairly easy to implement  this method in
{\it Mathematica\/}. The interested reader will find such an
implementation in Gallier \cite{Gallbook2}.
In the interest of brevity, we content ourselves with
some examples.

\bigskip\noindent
{\it Example\/} 1.
The algorithm is illustrated by the following example 
of an ellipsoid defined by the fractions
$$
x(u,v) = \frac{2c_1u}{u^2 + v^2 + 1},\quad
y(u,v) = \frac{2c_2v}{u^2 + v^2 + 1},\quad
z(u,v) = \frac{c_3(u^2 + v^2 - 1)}{u^2 + v^2 + 1}.
$$

\medskip
It is easily verified that this representation of the ellipsoid
is derived from the stereographic projection from the north pole
onto the plane $z = 0$. The coordinates of a  point on the
sphere are the coordinates of the
image of a point $(u, v)$ the $x O y$ plane,
under the inverse of stereographic projection.
We leave as an exercise to show that the following triangular control net
for $c_1 = 4$, $c_2 = 3$, $c_3= 2$, is obtained:

\medskip
\begin{verbatim}
net = {{0, 0, -2, 1}, {0, 3, -2, 1}, {0, 3, 0, 2},
       {4, 0, -2, 1}, {4, 3, -2, 1}, {4, 0, 0, 2}}
\end{verbatim}

\medskip
The following picture shows the result of iterating 
the subdivision algorithm $3$ times on the
nets {\tt net1\/} and  {\tt net2\/}:

\begin{figure}[H]
\centerline{
\includegraphics[width=3.5truein,height=3.5truein]{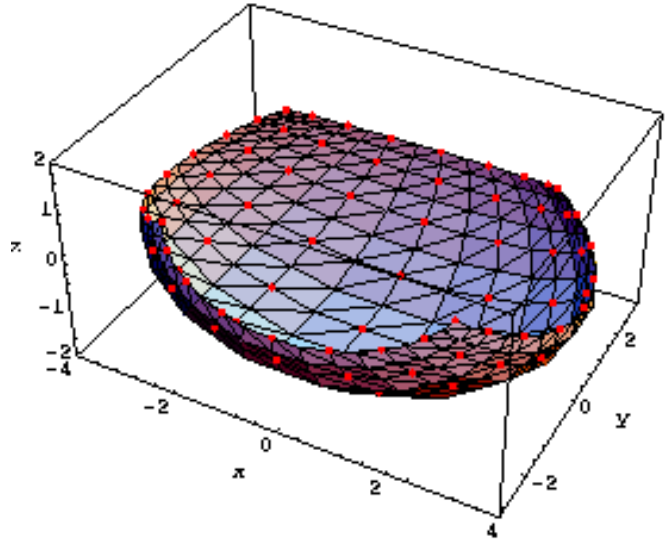}
}
\caption{Patches 1, 2, of an ellipsoid}
\end{figure}

Iterating the subdivision algorithm $3$ times on the
nets {\tt theta1\/} and  {\tt theta2\/} yields:

\begin{figure}[H]
\centerline{
\includegraphics[width=3.5truein]{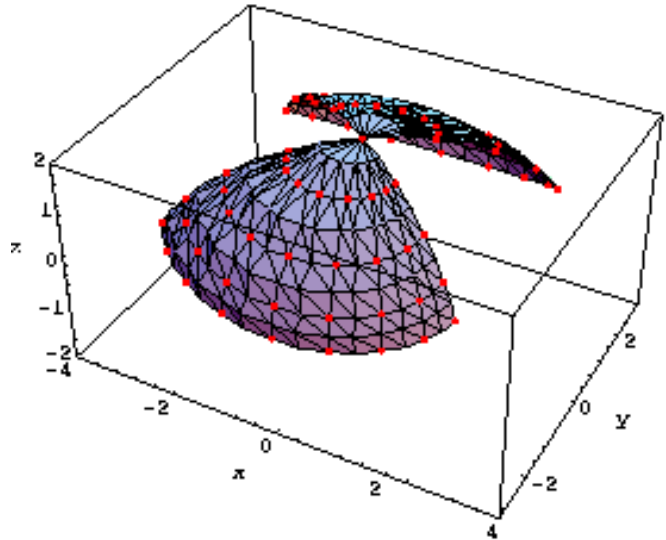}
}
\caption{Patches 3, 4, of an ellipsoid}
\end{figure}

Iterating the subdivision algorithm $3$ times on the
nets {\tt rho1\/} and  {\tt rho2\/} yields:

\begin{figure}[H]
\centerline{
\includegraphics[width=3truein]{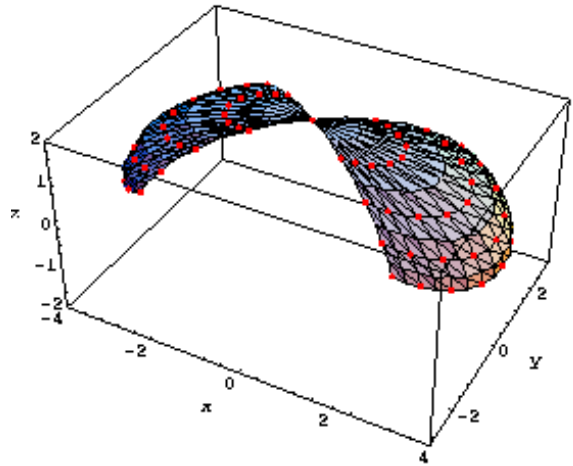}
}
\caption{Patches 5, 6, of an ellipsoid}
\end{figure}

The result of putting all these patches together is the entire
ellipsoid:

\begin{figure}[H]
\centerline{
\includegraphics[width=3.5truein]{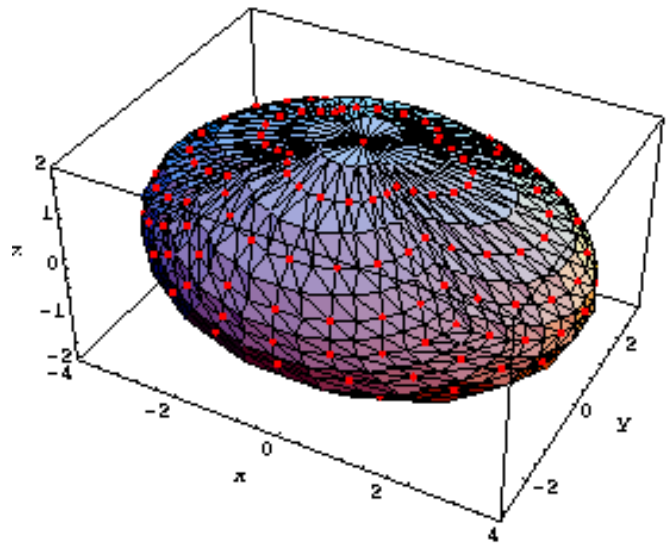}
}
\caption{An entire ellipsoid}
\end{figure}

Of course, we could have taken advantage of symmetries,
and our point is to illustrate the algorithm.

\medskip\noindent
{\it Example\/} 2.
The {\it Steiner roman surface\/} is the surface of implicit equation
$$x^2y^2 + y^2z^2 + x^2z^2 = 2xyz.$$
It is easily
verified that the following parameterization works:
$$
x(u,v) = \frac{2v}{u^2 + v^2 + 1},\quad
y(u,v) = \frac{2u}{u^2 + v^2 + 1},\quad
z(u,v) = \frac{2uv}{u^2 + v^2 + 1}.
$$

\begin{figure}[H]
\centerline{
\includegraphics[width=3truein]{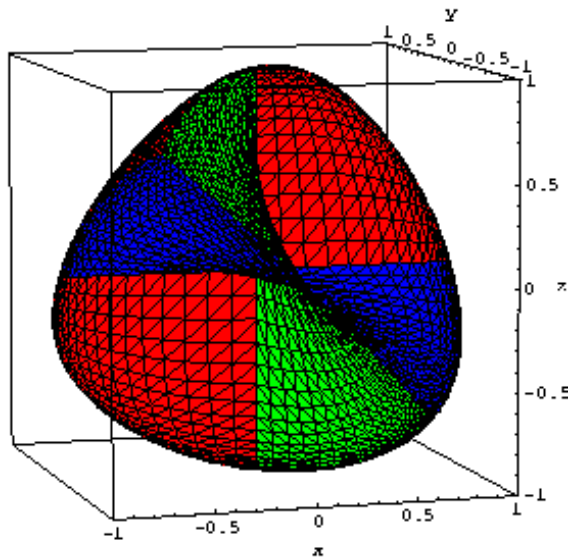}
}
\caption{The Steiner roman surface}
\label{steinfig1}
\end{figure}

It can be shown that this surface is contained inside
the tetrahedron defined by the planes 
$$ -x + y + z = 1,\quad
x - y + z = 1,\quad
 x + y - z = 1,\quad
-x  -y -z = 1,
$$
with $-1\leq x, y, z\leq 1$.
The surface touches these four planes along
ellipses, and at the middle of the six edges of the tetrahedron,
it has sharp edges. 
Furthermore, the surface is self-intersecting along the
axes, and is has four closed chambers.
A more extensive discussion can be found in Hilbert and
Cohn-Vossen \cite{Hilbert}, in particular, its relationship to
the heptahedron.
A triangular control net is easily obtained:

\begin{verbatim}
stein1 = {{0, 0, 0, 1}, {1, 0, 0, 1}, {1, 0, 0, 2}, 
       {0, 1, 0, 1}, {1, 1, 1, 1},  {0, 1, 0, 2}};
\end{verbatim}

We can display the entire surface using the method described in this section.
Indeed, all six patches are needed to obtain the entire surface.
One view of the surface obtained by subdividing $3$ times is shown below
(see Figure \ref{steinfig1}).
Patches 1 and 2 are colored blue, patches 3 and 4 are colored red,
and patches 5 and 6 are colored green. A closer look reveals that
the three colored patches are identical under
appropriate rigid motions, and fit perfectly.

\begin{figure}[H]
\centerline{
\includegraphics[width=3truein]{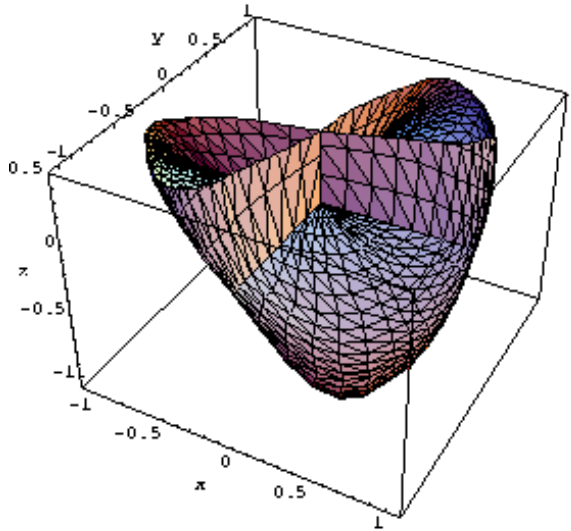}
}
\caption{A cut of the Steiner roman surface}
\label{steinfig2}
\end{figure}

Another revealing view (see Figure \ref{steinfig2})
is obtained by cutting off a top portion of
the surface. This way, it is clear that the surface has chambers.

\section{Splitting Triangular Rational Surfaces Into Four Triangular Patches}
\label{sec5}
As explained in Section \ref{sec1}, we obtain a partition
of the real projective plane $\rprospac{2}$ into four triangles
if we project an octahedron onto one of its faces from its center.
We sketch such a method, leaving the simple details to the reader.

\vskip 0.15cm
\begin{figure}[H]
%\begin{example}
  \begin{center}
    \begin{pspicture}(-2,-2)(5,5)
    \psline[linewidth=1.5pt](-2,0)(5,0)
    \psline[linewidth=1.5pt](0,-2)(0,5)
    \psline[linewidth=1.5pt](-2,5)(5,-2)
    \psdots[dotstyle=o,dotscale=1.5](3,0)
    \psdots[dotstyle=o,dotscale=1.5](0,3)
    \psdots[dotstyle=o,dotscale=1.5](0,0)
    \uput[90](3,0){$r$}
    \uput[0](0,3){$s$}
    \uput[45](0,0){$t$}
    \uput[45](0.6,0.6){$rst$}
    \uput[45](2,2){$T$}
    \uput[45](-1,-1){$T$}
    \uput[45](4.2,-0.9){$R$}
    \uput[45](-1.2,1.5){$R$}
    \uput[45](-0.7,4){$S$}    
    \uput[-90](1.5,-0.4){$S$}    
\end{pspicture}
  \end{center}
  \caption{Splitting $\rprospac{2}$ into four triangles}
%\end{example}
\end{figure}
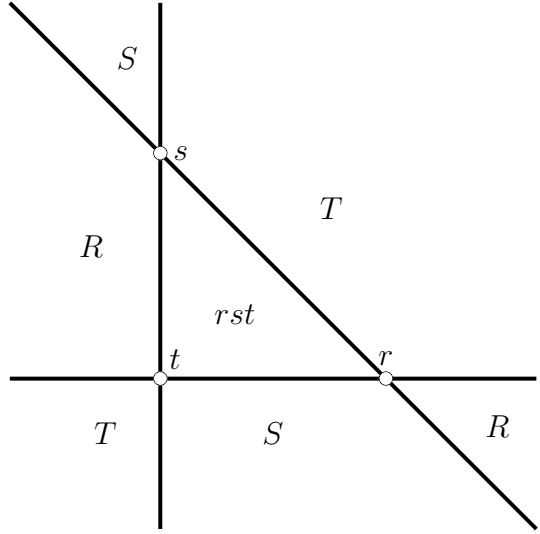

We let $r, s, t$ be the vertices of the central triangle.
The four triangles defined by the lines $\pairt{r}{s}$,  $\pairt{s}{t}$,
and  $\pairt{r}{t}$ are denoted as $rst$, $R$, $S$, and $T$,
where $R, S, T$ contain points at infinity.
It is easy to find three projectivities
$\mapdef{\varphi_i}{\rprospac{2}}{\rprospac{2}}$, $i = 1, 2, 3$,
such that $\varphi_1(rst) = R$,  $\varphi_2(rst) = S$, and
$\varphi_3(rst) = T$. Then, we get some rational surfaces
$F_i = F\circ \varphi_i$, $i = 1, 2, 3$. 
Indeed, if we 
use the model of $\rprospac{2}$ in $\reals^3$
where the points $r, s, t$ are considered as being in the plane $z = 1$,
it is immediately verified that the linear maps  
$$\eqaligneno{
(r, s, t) &\mapsto (-r, s, t),\cr
(r, s, t) &\mapsto (r, -s, t),\cr
(r, s, t) &\mapsto (r, s, -t),\cr
}$$
induce $\varphi_1, \varphi_2, \varphi_3$.
Furthermore, if the control net
(in $\hli{\affs}$) of the triangular surface $F$
w.r.t. the affine frame $\Delta rst$ is
$\alpha = (\alpha_{i,\,j,\,k})_{(i,j,k)\in\Delta_{\mdeg}}$,
it can be shown that the control nets $\theta^1, \theta^2$, 
and $\theta^3$ of the surfaces
$F_1, F_2, F_3$ w.r.t. $\Delta rts$ are given by the formulae
$$\eqaligneno{
\theta^{1}_{i,\, j,\, k} & = (-1)^{i} \alpha_{i,\, j,\, k},\cr
\theta^{2}_{i,\, j,\, k} & = (-1)^{j} \alpha_{i,\, j,\, k},\cr
\theta^{3}_{i,\, j,\, k} & = (-1)^{k} \alpha_{i,\, j,\, k}.\cr
}$$

\smallskip
Provided that there are no base points,
the traces of $F, F_1, F_2, F_3$ over $\Delta rst$
cover the entire trace of $F$ (over $\rprospac{2}$).
The upshot is that in order to draw a whole
rational surface
given by a triangular net $\alpha$  over $\Delta rst$, we simply have to
draw the four patches specified by
$\alpha$,  $\theta^1, \theta^2$, and $\theta^3$, over $\Delta rst$.

\medskip
For example, we can apply the above method to the 
Steiner roman surface specified by the triangular net
given in Example 2.
It turns out that the patch $F_3$ is quite distorted. Applying the method to
a net over a bigger triangle helps reduce the distorsion.
In particular, we can send $r$ and $s$ to infinity, in which case
the method ends up being equivalent to a method due to Bajaj and Royappa
\cite{Bajaj94a,Bajaj95a}. Their method is based on the observation that
the four maps 
$$(u, v)\mapsto \biggl(\frac{\sigma_1 u}{1 - u - v},\> 
\frac{\sigma_2 v}{1 - u - v}\biggr),$$
where $\sigma_i\in \{-1, 1\}$ for $i = 1, 2$, 
map the triangle $((1, 0), (0, 1), (0, 0))$  bijectively onto
the four quadrants of the plane respectively.
However, they do not consider the problem of computing
the control nets of the surfaces
$$F\biggl(\frac{\sigma_1 u}{1 - u - v},\> \frac{\sigma_2 v}{1 - u - v}\biggr).$$

\medskip
Another method for drawing triangular rational surfaces was also investigated
by DeRose \cite{deRose91} who credits
Patterson \cite{Patterson86} for the original idea behind
the method. Basically, the method consists in using the homogeneous
Bernstein polynomials 
$\binom{\mdeg}{i\> j\> k}\, u^i v^j w^k$, where $i + j + k = \mdeg$,
and to view a triangular rational surface
as a rational map from the real projective plane. Then,
by using any $3D$ model of the projective plane, it is possible to draw
whole rational surface in one piece. For example, DeRose
suggests to use an octahedron. However, the problem of finding
efficient ways of computing control points  is not addressed.

\section{Splitting Rectangular Rational Surfaces Into Four Rectangular Patches}
\label{sec6}
In this section, 
we show that every  rectangular rational surface can be obtained as the
union of four rectangular patches, and that the control nets for these
patches can be computed very easily from the original control net.
The idea is simple: we partition $\rprospac{1}\times \rprospac{1}$
into the four regions associated with the partitioning of
$\rprospac{1}$ into $[-1, 1]$ and
$\rprospac{1} - [-1, 1]$. 
Let $\varphi$ be the projectivity of $\rprospac{1}$ defined such that
$$\varphi(u, t) =(t, u).$$

We also define the following rectangular surfaces.
\begin{defin}
\label{fphispisdefsq}
{\em
Given an affine space $\affs$ of dimension $\geq 3$,
for every rectangular rational surface
$\mapdef{F}{\pcompl{\affreal}\times \pcompl{\affreal}}{\pcompl{\affs}}$
of bidegree $\pairt{\pdeg}{\qdeg}$ specified by 
some $\pairt{\pdeg}{\qdeg}$-symmetric multilinear map 
$\mapdef{f}{(\hli{\affreal})^{\pdeg}\times (\hli{\affreal})^{\qdeg}}{\hli{\affs}}$,
define the three
$\pairt{\pdeg}{\qdeg}$-symmetric multilinear maps 
$\mapdef{f_i}{(\hli{\affreal})^{\pdeg}\times (\hli{\affreal})^{\qdeg}}{\hli{\affs}}$,
$i = 1, 2, 3$, such that
$$\eqaligneno{
f_1((u_1, t_1),\ldots, (u_{\pdeg}, t_{\pdeg}),
(v_1, s_1),\ldots, (v_{\qdeg}, s_{\qdeg})) &=
f(\varphi(u_1, t_1),\ldots, \varphi(u_{\pdeg}, t_{\pdeg}),
(v_1, s_1),\ldots, (v_{\qdeg}, s_{\qdeg})),\cr
f_2((u_1, t_1),\ldots, (u_{\pdeg}, t_{\pdeg}),
(v_1, s_1),\ldots, (v_{\qdeg}, s_{\qdeg})) &=
f((u_1, t_1),\ldots, (u_{\pdeg}, t_{\pdeg}),
\varphi(v_1, s_1),\ldots, \varphi(v_{\qdeg}, s_{\qdeg})),\cr
f_3((u_1, t_1),\ldots, (u_{\pdeg}, t_{\pdeg}),
(v_1, s_1),\ldots, (v_{\qdeg}, s_{\qdeg})) &=
f(\varphi(u_1, t_1),\ldots, \varphi(u_{\pdeg}, t_{\pdeg}),
\varphi(v_1, s_1),\ldots, \varphi(v_{\qdeg}, s_{\qdeg})).\cr
}$$
}
\end{defin}

\medskip
The following lemma shows that 
provided that there are no base points, a rectangular rational surface
is the union of four rectangular patches, and that given
a rectangular net $\alpha$ w.r.t. $(-1, 1)\times (-1, 1)$,
the other three nets can be obtained very easily from $\alpha$.

\begin{lemma}
\label{rendlem2sq}
Given an affine space $\affs$ of dimension $\geq 3$,
for every rectangular rational surface
$\mapdef{F}{\pcompl{\affreal}\times \pcompl{\affreal}}{\pcompl{\affs}}$
of bidegree $\pairt{\pdeg}{\qdeg}$ specified by 
some $\pairt{\pdeg}{\qdeg}$-symmetric multilinear map 
$\mapdef{f}{(\hli{\affreal})^{\pdeg}\times (\hli{\affreal})^{\qdeg}}{\hli{\affs}}$,
if $f_1, f_2, f_3$ 
are the $\pairt{\pdeg}{\qdeg}$-symmetric multilinear maps of definition 
\ref{fphispisdefsq}, 
except for the base points (if any),
the trace $F_{1}([-1, 1]\times [-1, 1])$
is the trace of $F$  over $\varphi([-1, 1])\times [-1, 1]$,
the trace $F_{2}([-1, 1]\times [-1, 1])$
is the trace of $F$  over $[-1, 1]\times \varphi([-1, 1])$,
and the trace $F_{3}([-1, 1]\times [-1, 1])$ 
is the trace of $F$  over $\varphi([-1, 1]) \times \varphi([-1, 1])$.
Furthermore, if the control net 
(in $\hli{\affs}$) of the rectangular surface $F$
w.r.t.  $(-1, 1)\times (-1, 1)$ is
$$\alpha = (\alpha_{i,\,j})_{0\leq i \leq\pdeg,\, 0\leq j \leq \qdeg},$$
the control nets $\theta^{1}$, $\theta^{2}$, and  $\theta^{3}$
(in $\hli{\affs}$) of the rectangular surfaces $F_{1}, F_{2}, F_{3}$ 
w.r.t.  $(-1, 1)\times (-1, 1)$ is
are given by the equations
$$\eqaligneno{
\theta_{i,\, j}^{1} &= (-1)^{\pdeg - i}\> \alpha_{i,\, j},\cr
\theta_{i,\, j}^{2} &= (-1)^{\qdeg - j}\> \alpha_{i,\, j},\cr
\theta_{i,\, j}^{3} &= (-1)^{\pdeg + \qdeg - i - j}\> \alpha_{i,\, j}.\cr
}$$
\end{lemma}

The proof is quite simple and left as an exercise.
Actually, the same result applies to surfaces specified by
a rectangular net over $[r_1, s_1]\times [r_2, s_2]$
for any affine frames $(r_1, s_1)$ and $(r_2, s_2)$,  since
we can use the projectivity
$$\varphi(t) = \frac{(s + r)t - 2rs}{2t - (s + r)}$$
that maps $[r, s]$ onto $\rprospac{1} - ]r, s[$.
The upshot is that in order to draw a whole rational surface
specified by a rectangular net $\alpha$ w.r.t.
$(r_1, s_1)\times (r_2, s_2)$,
we simply have to compute the nets $\theta^1, \theta^2, \theta^3$, 
which is very cheap, and draw the corresponding rectangular patches.
For example, a torus can be defined by the following rectangular net
of bidegree $\pairt{2}{2}$ w.r.t.  $(-1, 1)\times (-1, 1)$:

\begin{verbatim}
tornet4 = {{0, -(a + b), 0, 4}, {0, 0, 4c, 0}, {0, (-a + b), 0, 4},
   {4(a + b), 0, 0, 0}, {0, 0, 0, 0}, {4(a - b), 0, 0, 0},
   {0, a + b, 0, 4}, {0, 0, 4c, 0}, {0, a - b, 0, 4}}
\end{verbatim}

For $a = 2$, $b = 1$, $c = 1$, we get

\begin{verbatim}
tornet4 = {{0, -3, 0, 4}, {0, 0, 4, 0}, {0, -1, 0, 4},
   {12, 0, 0, 0}, {0, 0, 0, 0}, {4, 0, 0, 0},
   {0, 3, 0, 4}, {0, 0, 4, 0}, {0, 1, 0, 4}}
\end{verbatim}

The result of subdividing the patches associated with $F$, $F_1$, $F_2$ and $F_3$
is shown below.

\begin{figure}[H]
\centerline{
\includegraphics[width=3.5truein]{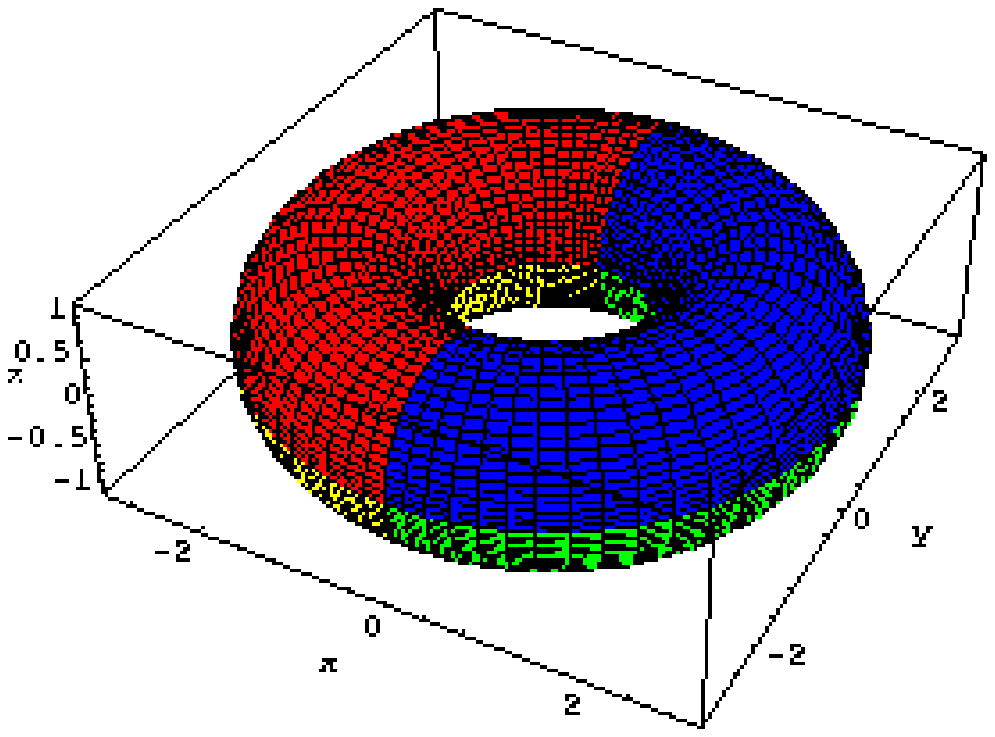}
}
\caption{A torus}
\end{figure}

\medskip
On the other hand, the method applied to a rectangular net
of bidegree $\pairt{2}{2}$ for an ellipsoid
yields base points. For example, it can be shown that
a control net of bidegree $\pairt{2}{2}$ w.r.t.
$(-1, 1)\times (-1, 1)$ for an ellipsoid is given by:

\begin{verbatim}
recelnet3 =  {{-8/3, -2, 2/3, 3}, {-8, 0, -2, 1}, {-8/3, 2, 2/3, 3}, 
   {0, -6, -2, 1}, {0, 0, 6, -1}, {0, 6, -2, 1}, 
   {8/3, -2, 2/3, 3}, {8, 0, -2, 1},  {8/3, 2, 2/3, 3}}
\end{verbatim}

Unfortunately, the patch corresponding to $F_3$ has a base point.
The same thing happens for the Steiner roman surface.
This is not surprising since neither the sphere nor the
real projective  plane are of the same topological type as the torus.
It can be shown that a control net of bidegree 
$\pairt{2}{2}$ w.r.t.
$(-1, 1)\times (-1, 1)$ for the Steiner roman surface is given by:

\begin{verbatim}
sqstein3 = {{-2/3, -2/3, 2/3, 3}, {0, -2, 0, 1}, {2/3, -2/3, -2/3, 3}, 
   {-2, 0, 0, 1}, {0, 0, 0, -1}, {2, 0, 0, 1}, 
   {-2/3, 2/3, -2/3, 3}, {0, 2, 0, 1}, {2/3, 2/3, 2/3, 3}};
\end{verbatim}

Again, the the patch corresponding to $F_3$  has base points.

\medskip
Another method for drawing rectangular rational surfaces was  investigated
by DeRose \cite{deRose91} who credits
Patterson \cite{Patterson86} for the original idea behind
the method. Basically, the method consists in using the homogeneous
Bernstein polynomials 
$\binom{\pdeg}{i}\binom{\qdeg}{j}\, u^it_1^{\pdeg - i} v^j t_2^{\qdeg - j}$, 
and to view a rectangular rational surface
as a rational map from $\rprospac{1}\times\rprospac{1}$. Then,
by using any $2D$ model of the projective line, it is possible to draw
a whole rational surface in one piece.

\medskip
In general, it is not easy to remove base points. This involves
a technique from algebraic geometry known as ``blowing-up'' 
(see Fulton \cite{Fulton} or Harris \cite{Harris}).
In the next section, we will present a method
for resolving base points in the case of
triangular rational surfaces. However, we have not worked out
the resolution of base points in the
case of rectangular rational surfaces. 
We leave this problem as an interesting challenge to the reader.

\section{Resolving Base Points}
\label{sec7}
We now consider the case in which $F_{\varphi}$ and
$F_{\psi}$ (as defined in Section \ref{sec4})  have base points. 
An example for which this happens is the torus. 

\medskip\noindent
{\it Example\/} 3.
An elliptic torus can be defined parametrically 
as follows:
$$\eqaligneno{
x &= (a - b\>\sin \varphi)\cos\theta,\cr
y &= (a - b\>\sin \varphi)\sin\theta,\cr
z &= c\>\cos\varphi.\cr
}$$

\medskip
As usual, we obtain a rational parameterization by expressing
$\cos t$ and $\sin t$ in terms of $\tan(t/2)$, and we get 
the fractions
$$\eqaligneno{
x &= \frac{(1 - u^2)(a(1 + v^2) - 2bv)}{(1 + u^2)(1 + v^2)},\cr
y &= \frac{2u(a(1 + v^2) - 2bv)}{(1 + u^2)(1 + v^2)},\cr
z &= \frac{c(1 - v^2)}{1 + v^2}.\cr
}$$

Thus, the torus as a rational surface $F$ is defined by
$$\eqaligneno{
x(u, v) &= (1 - u^2)(a(1 + v^2) - 2bv),\cr
y(u, v) &= 2u(a(1 + v^2) - 2bv),\cr
z(u, v) &= c(1 + u^2)(1 - v^2),\cr
w(u, v) &= (1 + u^2)(1 + v^2).\cr
}$$

\medskip
Rendering $F$ over $[-1, 1]\times [-1, 1]$ yields one fourth of
the torus, specifically, the front half of the upper half.
Performing the change of variables
$$(u, v) \mapsto \biggl(\frac{v}{u}, \frac{1}{u}\biggr),$$
the rational surface $F_{\varphi}$ is defined by
$$\eqaligneno{
x(u, v) &= (u^2 - v^2)(a(1 + u^2) - 2bu),\cr
y(u, v) &= 2uv(a(1 + u^2) - 2bu),\cr
z(u, v) &= c(u^2 - 1)(u^2 + v^2),\cr
w(u, v) &= (u^2 + v^2)(u^2 + 1).\cr
}$$

Unfortunately, $x(0, 0) = y(0, 0) = z(0, 0) = w(0, 0) = 0$, 
and $(0, 0)$ is a base point of $F_{\varphi}$.

\medskip
Performing the change of variables
$$(u, v) \mapsto \biggl(\frac{1}{v}, \frac{u}{v}\biggr),$$
the rational surface $F_{\psi}$ is defined by
$$\eqaligneno{
x(u, v) &= (v^2 - 1)(a(u^2 + v^2) - 2buv),\cr
y(u, v) &= 2v(a(u^2 + v^2) - 2buv),\cr
z(u, v) &= c(v^2 - u^2)(v^2 + 1),\cr
w(u, v) &= (u^2 + v^2)(v^2 + 1).\cr
}$$

Unfortunately, we also have $x(0, 0) = y(0, 0) = z(0, 0) = w(0, 0) = 0$,
and $(0, 0)$ is a base point of $F_{\psi}$.

\medskip
If we try to render the rational surfaces $F_{\varphi}$ and $F_{\psi}$
over $[-1, 1]\times [-1, 1]$, we discover that some regions
of these surfaces are not drawn properly. In these regions,
there are holes and many lines segments shooting in all directions!
The problem is that $(0, 0)$ is
a discontinuity point for both surfaces, and that the limit
reached when $u$ and $v$ approach $0$ depends very much on the
ratio $v/u$. One way to understand what happens is to let
$v = ku$,  simplify the fractions, and see what is the limit when
$u$ approaches $0$.
For $F_{\varphi}$, 
after calculations, we find that the limit when $u$ approaches $0$ is
$$\biggl(\frac{a(1 - k^2)}{1 + k^2},\> 
\frac{2ak}{1 + k^2},\> -c\biggr),$$
which corresponds to the circle of radius $a$ in the plane
$z = -c$.
For $F_{\psi}$, 
after calculations, we find that the limit when $u$ approaches $0$ is
$$\biggl(-a + \frac{2bk}{1 + k^2},\> 0,\>
-\frac{c(1 - k^2)}{1 + k^2}\biggr),$$
which corresponds to an ellipse in the plane $y = 0$, centered at the point
$(-a, 0, 0)$.
It is indeed in  the neighborhood of these two curves on the torus
that  $F_{\varphi}$ and $F_{\psi}$ are not drawn properly.

\medskip
We now propose  a method to resolve the singularities caused by
base points. The method is
inspired by a technique in algebraic geometry known as 
``blowing-up'' (see Fulton \cite{Fulton} or Harris \cite{Harris}).
What is new is that we give formulae for
computing ``resolved'' control nets.

\medskip
In most cases, base points occur during
a subdivision step in which a triangular net with
a corner of zeros appears. Using a change of base triangle if 
necessary, it can be assumed without loss of generality that
the corner of zeros has $t$ as one of its vertices.
If we display control nets (in $\hli{\affs}$) with $F(r)$ at the top corner,
$F(s)$ as the rightmost lower corner, and $F(t)$ as the
leftmost lower corner, a control net 
$\theta = (\theta_{i,\,j,\,k})_{(i,j,k)\in\Delta_{\mdeg}}$
of degree $\mdeg$
has the following shape:
$$\displaylignes{
\hfill \times\hfill\cr
\hfill \times \times\hfill\cr
\hfill \ldots \hfill\cr
\hfill \times\> \times\> \ldots\times\> \> \times\hfill\cr
\hfill \vectorr{0}\>  
\times\> \times\> \ldots\times\> \> \times\hfill\cr
\hfill \vectorr{0}\> \> \vectorr{0}\> 
\times\> \times\> \ldots\times\> \> \times\hfill\cr
\hfill \ldots \hfill\cr
\hfill \vectorr{0}\>  \ldots \> \vectorr{0}\> 
\times\> \times\> \ldots\times\> \> \times\hfill\cr
\hfill \underbrace{\vectorr{0}\> \vectorr{0}\> 
\ldots \>\vectorr{0}\> \vectorr{0}}_{\ndeg}\> 
\underbrace{\times\> \times\> \ldots
\times\> \> \times}_{\mdeg + 1 - \ndeg}\hfill\cr
}$$

\medskip
It is assumed that all entries designated as $\times$ are
nonzero. 
The more general case can be treated, but it is 
computationally too expensive to be practical.

\medskip
Given an affine frame $\Delta rst$ in the plane,
recall that a rational surface $F$ of degree $\mdeg$ defined by
the control net
$\theta = (\theta_{i,\,j,\,k})_{(i,j,k)\in\Delta_{\mdeg}}$
is the projection onto $\pcompl{\affs}$ of the
polynomial surface $G$ in $\hli{\affs}$ defined by $\theta$.
Also, we have 
$$G(u, v) = \sum_{i + j + k = \mdeg} 
\theta_{i,\, j,\, k}\, \frac{\mdeg !}{i!j!k!}\, u^{i}v^{j}(1 - u - v)^{k},$$
for all $u, v\in \reals$.
It will be convenient to assume that if  $\theta_{i,\, j,\, k}\in\hli{\affs}$
is a weighted point, then its weight is denoted as $w_{i, j, k}$,
and if $\theta_{i,\, j,\, k}$ is a control vector, then we assign it
the weight $w_{i, j, k} = 0$.
If we define $w(u, v)$ as
$$w(u, v) = \sum_{i + j + k = \mdeg} 
w_{i, j, k}\, \frac{\mdeg !}{i!j!k!}\, u^{i}v^{j}(1 - u - v)^{k},$$
whenever $w(u, v) \not= 0$, we have
$$F(u, v) = \sum_{i + j + k = \mdeg} 
\theta_{i,\, j,\, k}\, \frac{\mdeg !}{i!j!k!}\, 
\frac{u^{i}v^{j}(1 - u - v)^{k}}{w(u, v)},$$
for all $u, v\in \reals$.

\medskip
The ``blowing-up'' method used here relies on the following observation
based on an idea of Warren \cite{Warren92}.
Given the polynomial surface $G$ in $\affs$ (and $w$), we define
the polynomial surface $G_{b}$ and $w_{b}$
as follows:
$$\eqaligneno{
G_{b}(\alpha, \beta) &= G(\alpha(1 - \beta), \alpha\beta),\cr
w_{b}(\alpha, \beta) &= w(\alpha(1 - \beta), \alpha\beta).\cr
}$$
Since $\alpha(1 - \beta) + \alpha\beta = \alpha$,
we get
$$G_{b}(\alpha, \beta) = 
\sum_{i + j + k = \mdeg} 
\theta_{i,\, j,\, k}\, \frac{\mdeg !}{i!j!k!}\, 
\alpha^{i + j}(1 - \alpha)^{k}\beta^{j}(1 - \beta)^{i},$$
and
$$w_{b}(\alpha, \beta)  = \sum_{i + j + k = \mdeg} 
w_{i, j, k}\, \frac{\mdeg !}{i!j!k!}\, 
\alpha^{i + j}(1 - \alpha)^{k}\beta^{j}(1 - \beta)^{i}.$$
Now, if $\theta_{i,\, j,\, k} = \vectorr{0}$ for $i + j < \ndeg$
(with $i + j + k = \mdeg$), we note that both
$G_{b}(\alpha, \beta)$ and 
$w_{b}(\alpha, \beta)$ are divisible by  $\alpha^{\ndeg}$.
If we define the polynomial surface $\widetilde{G}$ 
(and $\widetilde{w}$), such that
$$\widetilde{G}(\alpha, \beta) = 
\frac{G_{b}(\alpha, \beta)}{\alpha^{\ndeg}}$$
 and
$$\widetilde{w}(\alpha, \beta) = 
\frac{w_{b}(\alpha, \beta)}{\alpha^{\ndeg}},$$
then we have
$$\frac{G_{b}(\alpha, \beta)}{w_{b}(\alpha, \beta)}
= \frac{\widetilde{G}(\alpha, \beta)}{\widetilde{w}(\alpha, \beta)},$$
for all $\alpha\not= 0$.
Furthermore when $\alpha = 0$, we have
$$\widetilde{G}(0, \beta) = 
\sum_{i + j = \ndeg} 
\theta_{i,\, j,\, \mdeg - \ndeg}\, \frac{\mdeg !}{i!j!(\mdeg - \ndeg)!}\, 
\beta^{j}(1 - \beta)^{i},$$
and
$$\widetilde{w}(0, \beta)  = \sum_{i + j = \ndeg} 
w_{i, j, \mdeg - \ndeg}\, \frac{\mdeg !}{i!j!(\mdeg - \ndeg)!}\, 
\beta^{j}(1 - \beta)^{i}.$$
Thus, for all $\beta$ for which $\widetilde{G}(0, \beta)$
and $\widetilde{w}(0, \beta)$ are not simultaneously null,
$$\frac{\widetilde{G}(0, \beta)}{\widetilde{w}(0, \beta)}$$
is defined, and the polynomial surface $\widetilde{G}$
defines the rational surface $\widetilde{F}$ such that
$$\widetilde{F}(\alpha, \beta) = 
\frac{\widetilde{G}(\alpha, \beta)}{\widetilde{w}(\alpha, \beta)}.$$

\medskip
Thus, what happens is that the triangular patch $F$ over $\Delta rst$ is
really a four-sided patch, the point $F(t)$ being ``blown up''
into the rational curve of degree $n$ whose
control points are 
$$(\theta_{i,\, j,\, \mdeg - \ndeg})_{i + j = \ndeg}.$$
If this rational curve has no base points, then
the rational surface patch $\widetilde{F}$
defined by the polynomial surface $\widetilde{G}$
has no base point, and it extends
the surface patch $F$ over $\Delta rst$.
If it has base points, they are common zeros of some polynomials
in $\beta$, and by simplifying by common factors and using
continuity, we could eliminate these base points.
For simplicity, we will assume that the boundary curve has
no base points.

\medskip
Viewing  $\widetilde{G}$  as a bipolynomial surface, note
that $\widetilde{G}$ has bidegree $\pairt{\mdeg - \ndeg}{\mdeg}$.
Also observe that the function
$$(\alpha, \beta) \mapsto (\alpha(1 - \beta), \alpha\beta)$$
maps the unit square with vertices  
$$(0, 0), (0, 1), (1, 1), (1, 0)$$
onto the triangle $\Delta rst = ((0, 1), (1, 0), (0, 0))$,
in such a way that the edge $((0, 0), (0, 1))$ is mapped
onto $t$, the vertex $(1, 1)$ is mapped onto $s$, and
the vertex $(1, 0)$ is mapped onto $r$. Furthermore, if
$u = \alpha(1 - \beta)$ and $v = \alpha\beta$,
we get
$$\alpha = u + v$$
and
$$\beta = \frac{v}{u + v}$$
and thus, the map is invertible except on the line
$u + v = 0$. Thus, we can think of the inverse map as
``blowing up'' the affine frame $\Delta rst$ into the
unit square. Specifically, the point $t$ is ``blown up'' 
into the edge $((0, 0), (0, 1))$.

\medskip

\begin{figure}[H]
%\begin{example}
  \begin{center}
    \begin{pspicture}(0,-0.5)(9,3)
    \psline[linewidth=1.5pt](0,0)(0,3)
    \psline[linewidth=1pt](0,0)(3,0)
    \psline[linewidth=1pt](3,0)(3,3)
    \psline[linewidth=1pt](0,3)(3,3)
    \psline[linewidth=1pt](6,0)(6,3)
    \psline[linewidth=1pt](6,0)(9,0)
    \psline[linewidth=1pt](6,3)(9,0)
    \uput[-135](0,0){$(0, 0)$}
    \uput[135](0,3){$(0, 1)$}
    \uput[-45](3,0){$(1, 0)$}
    \uput[45](3,3){$(1, 1)$}
    \uput[-135](6,0){$t$}
    \uput[45](6,3){$s$}
    \uput[-45](9,0){$r$}
    \end{pspicture}
  \end{center}
  \caption{Blowing up a triangle into a square}
%\end{example}
\end{figure}
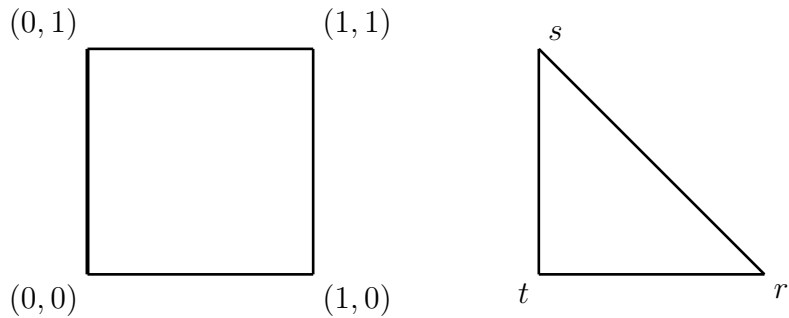

The only remaining problem is that the above method yields
a rational square patch not given by a control net,
and that we often need to map the unit square to an
arbitrary base triangle. The second problem is easily
solved. Assume that the affine frame
$\Delta rst$ has coordinates $((r_1, r_2), (s_1, s_2), (t_1, t_2))$.
It is easily seen that the map defined such that
$$\eqaligneno{
u &= (s_1 - r_1)\alpha\beta + (r_1 - t_1)\beta + t_1,\cr
v &= (s_2 - r_2)\alpha\beta + (r_2 - t_2)\beta + t_2,\cr
}$$
maps the unit square to the triangle $\Delta rst$, in such a way
that the edge $((0, 0), (0, 1))$ is mapped
onto $t$, the vertex $(1, 1)$ is mapped onto $s$, and
the vertex $(1, 0)$ is mapped onto $r$. 
Some simple calculations show that
$$\eqaligneno{
\alpha &= \frac{(s_2 - r_2)(u - t_1) - (s_1 - r_1)(v - t_2)}
{(r_1 - t_1)(s_2 -r_2) - (r_2 - t_2)(s_1 - r_1)},\cr
\beta &= \frac{(r_1 - t_1)(v - t_2) - (r_2 - t_2)(u - t_1)}
{(s_2 - r_2)(u - t_1) - (s_1 - r_1)(v - t_2)},\cr
}$$
and thus, the map is only invertible outside the line of equation
$$(s_2 - r_2)(u - t_1) - (s_1 - r_1)(v - t_2) = 0,$$
the parallel to the vector $(s_1 - r_1, s_2 - r_2)$ through $t$.

\medskip
Now, if $\mapdef{g}{(\s{P})^{\mdeg}}{\hli{\affs}}$
is the polar form associated with $G$, 
we can compute the polar form 
$\mapdef{g_{b}}{(\s{P})^{\mdeg}
\times (\s{P})^{\mdeg}}{\hli{\affs}}$
associated with the bipolynomial surface $G_{b}$ 
as follows:
$$g_{b}(u_1,\ldots, u_{m}, v_1,\ldots, v_{m}) =
 \frac{1}{\mdeg !}\> \sum_{\sigma\in \s{S}_{\mdeg}}
g((u_{1}(1 - v_{\sigma(1)}), u_{1}v_{\sigma(1)}),\ldots, 
(u_{\mdeg}(1 - v_{\sigma(\mdeg)}), u_{\mdeg}v_{\sigma(\mdeg)})),$$
where $\s{S}_{\mdeg}$ denotes the group of permutations on
$\{1, \ldots, \mdeg\}$. 
The above formula corresponds to the case of the simple mapping
$u = \alpha(1 - \beta)$, $v = \alpha\beta$,
and it is obvious how to adapt it to the more general
map
$$\eqaligneno{
u &= (s_1 - r_1)\alpha\beta + (r_1 - t_1)\beta + t_1,\cr
v &= (s_2 - r_2)\alpha\beta + (r_2 - t_2)\beta + t_2,\cr
}$$

\medskip
Now, over the affine basis $(0, 1)$, the square control net
$\theta{\square} = 
(\theta{\square}_{i, j})_{0\leq i, j \leq \mdeg}$ associated with 
$g_{b}$ is defined such that
$$\theta{\square}_{i, j} = 
g_{b}(\underbrace{0,\ldots, 0}_{\mdeg - i},
\underbrace{1,\ldots, 1}_{i},
\underbrace{0,\ldots, 0}_{\mdeg - j},
\underbrace{1,\ldots, 1}_{j}).$$
However, if $\theta_{i,\, j,\, k} = 0$ for
$i + j < \ndeg$, with $i + j + k= \mdeg$,
then $\theta{\square}_{i, j} = 0$ for $i < \ndeg$, and thus
we obtain the rectangular net $\widetilde{\theta}{\square}$
of degree $(\mdeg - \ndeg, \mdeg)$ associated with $\widetilde{g}$, given by
$$\widetilde{\theta}{\square} = 
(\theta{\square}_{i, j})_{\ndeg \leq i \leq \mdeg,\, 0\leq j \leq \mdeg},$$
which corresponds to the rational surface defined by
$\widetilde{G}$.

\medskip
Thus, we know how to compute a rectangular net for the blown-up
version $\widetilde{G}$ of $G$. 
A triangular net of degree $2\mdeg - \ndeg$ can easily 
be obtained. Indeed, there is a simple way for converting
the polar form
$\mapdef{g}{(\s{P})^{\pdeg}\times (\s{P})^{\qdeg}}{\hli{\affs}}$
of a bipolynomial surface of degree $(\pdeg, \qdeg)$ into
a symmetric multilinear polar form
$\mapdef{g_{\Delta}}{(\s{P})^{\pdeg + \qdeg}}{\hli{\affs}}$, 
using the following
formula: letting $\mdeg = \pdeg + \qdeg$, we have
$$g_{\Delta}((u_{1}, v_{1}),\ldots, (u_{\mdeg}, v_{\mdeg})) =
\frac{1}{\binom{\mdeg}{\pdeg}}
%\sum_{\scriptstyle I \cup J = \{1,\ldots, \mdeg\}\atop
%       {\scriptstyle I \cap J = \emptyset\atop
%        \scriptstyle |I| = \pdeg,\, |J| = \qdeg
%       }
\sum_{
\begin{subarray}{c}
I \cup J = \{1,\ldots, \mdeg\}\\
    I \cap J = \emptyset\\
    |I| = \pdeg,\, |J| = \qdeg
\end{subarray}
     }
g(\prod_{i\in I} u_i, \prod_{j\in J} v_j),$$
where 
$$g(\prod_{i\in I} u_i, \prod_{j\in J} v_j) = 
g(u_{i_{1}},\ldots, u_{i_{\pdeg}}, v_{j_{1}},\ldots, v_{j_{\qdeg}}),$$
with $I = \{i_{1}, \ldots, i_{\pdeg}\}$, and
$J = \{j_{1}, \ldots, j_{\qdeg}\}$.

\medskip
Note that it is also possible to convert the polar form 
$\mapdef{f_{\Delta}}{(\s{P})^{\mdeg}}{\hli{\affs}}$
of a surface of degree $\mdeg$ into a symmetric 
$(\mdeg, \mdeg)$-multilinear polar form  
$\mapdef{g}{(\s{P})^{\mdeg}\times (\s{P})^{\mdeg}}{\hli{\affs}}$,
using the following formula: 
$$g(u_{1},\ldots,u_{\mdeg}, v_{1},\ldots,v_{\mdeg}) =
\frac{1}{\mdeg !}
\sum_{\sigma\in\s{S}_{\mdeg}}
f((u_{1}, v_{\sigma(1)}),\ldots, (u_{\mdeg}, v_{\sigma(\mdeg)})),$$
where $\s{S}_{\mdeg}$ denotes the group of permutations on
$\{1,\ldots, \mdeg\}$.

\medskip
Thus, we have a method for blowing up a control net $\theta$
of degree $\mdeg$ with a corner of zeros of size $\ndeg$, into a triangular
net $\widetilde{\theta}$ 
of degree $2\mdeg - \ndeg$, by first blowing up the triangular net
$\theta$ into a rectangular net $\widetilde{\theta}{\square}$,
and then converting $\widetilde{\theta}{\square}$ into a triangular net
$\widetilde{\theta}$.

\medskip
Again, it is fairly easy to implement the above method
in {\it Mathematica\/}
(see Gallier \cite{Gallbook2}).
In the interest of brevity, we content ouselves with
some examples.

\medskip
Going back to Example 3 of this section,
a torus, it turns out that in subdividing the nets {\tt theta1\/},
{\tt theta2\/}, {\tt rho1\/}, and {\tt rho2\/}, degenerate nets
with a corner of zeros are encountered. In fact, these corners
have two rows of zeros. and thus, the blowing up method
yields nets of degree $6$.
For example, the net corresponding to  {\tt theta1\/}
is resolved to a triangular net, which after $3$ iterations
of subdivision, yields the following picture:

\begin{figure}[H]
\centerline{
\includegraphics[width=3.5truein]{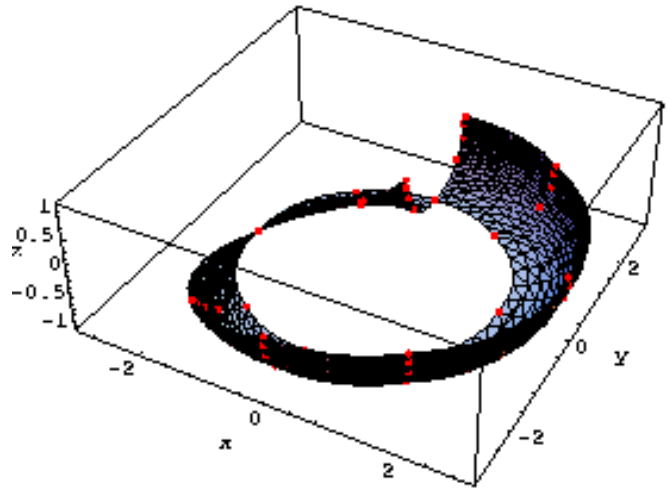}
}
\caption{A blow up of patch 3 of a torus}
\end{figure}

The net corresponding to  {\tt theta2\/}
is resolved to a triangular net, which after $3$ iterations
of subdivision, yields the following picture:

\begin{figure}[H]
\centerline{
\includegraphics[width=3.5truein]{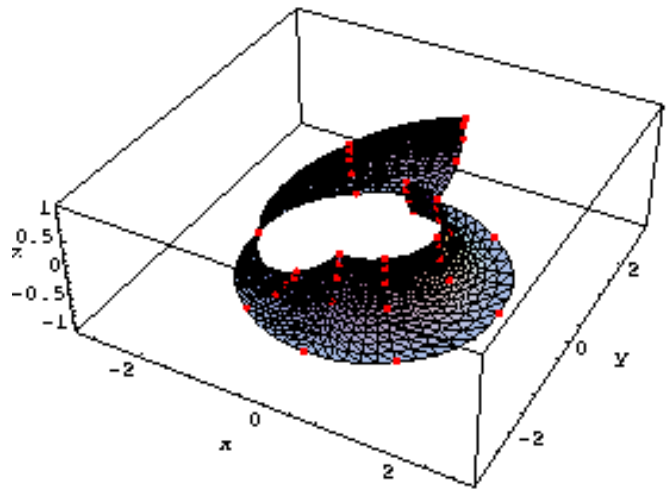}
}
\caption{A blow up of patch 4 of a torus}
\end{figure}

Displaying these two pictures together, we get a
shape reminicent of a horse-shoe crab!

\begin{figure}[H]
\centerline{
\includegraphics[width=3.2truein]{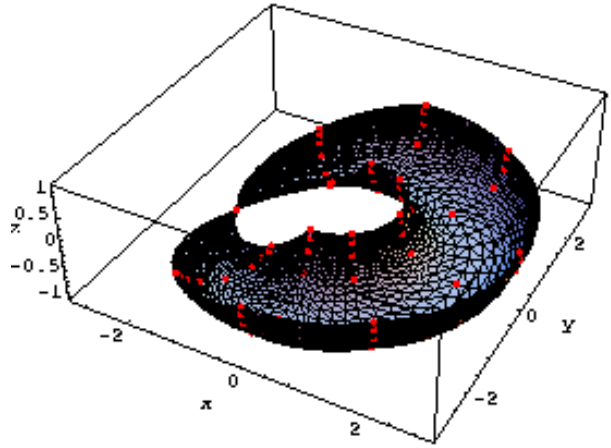}
}
\caption{A blow up of patches 4,3 of a torus}
\end{figure}

Similarly, blowing up the nets {\tt rho1\/}
and {\tt rho2\/} yields:

\begin{figure}[H]
\centerline{
\includegraphics[width=3.5truein]{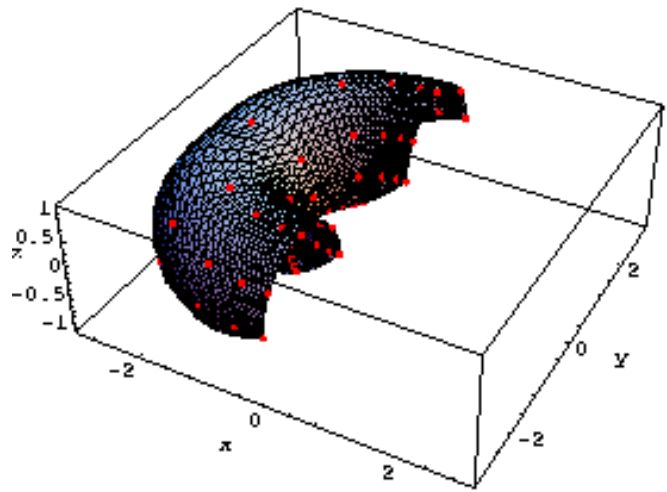}
}
\caption{A blow up of patches 5,6 of a torus}
\end{figure}

\medskip
Together with the two patches associated with
the square $[-1, 1]\times [-1, 1]$,
we get the entire torus.
Of course, we could have taken advantage of symmetries,
and our point is to illustrate the algorithm.

\section{Conclusion}
\label{sec8}
We have presented several  methods for drawing whole rational surfaces,
defined parametrically or in terms of weighted control points.
These methods rely on simple regular subdivisions of the real projective plane
or of the torus, and on  versions of
the de Casteljau algorithm. The main novelty is that
the new control nets are obtained very cheaply 
from the original control net by sign flipping
and permutation of indices.
One of the advantages of our method
is that it is incremental.
Indeed, the algorithm produces an approximation of the surface
as a sequence of control nets. Thus, if we wish to
get better accuracy, we can subdivide each control net in the list.
We can also achieve a zooming effect by selectively subdividing some subsequences
of control nets. Bajaj and Royappa 
\cite{Bajaj94a,Bajaj95a}
and DeRose \cite{deRose91}  
have also investigated methods
for drawing whole rational surfaces.
However, none of these papers address the problem
of computing control nets.
A weakness of our method is that it only applies to rational surfaces.
On the other hand, although restricted to rational surfaces,
our method is  efficient, at least when there are no base points.
We have also proposed a new method for
resolving base points, by computing some refined control nets.
Unfortunately, the present version of the method is exponential,
and not practical
as soon as the degree becomes greater than $4$. Part of the problem is
that our method first computes a rectangular control net which is then
converted to a triangular net, and this conversion process
is exponential. It would be interesting to compute directly
a triangular net.

\bigskip
{\it Acknowledgement\/}: We wish to thank Doug de Carlo for some 
very helpful comments.
\bibliography{cadgeom.bib}
\bibliographystyle{plain} 

\end{document}